\documentclass[PAPER, atlasdraft=false, cernpreprint,UKenglish, texmf, orcidlogo]{atlasdoc}
\usepackage{atlaspackage,subfig}
\usepackage{atlasbiblatex}

\usepackage{atlasphysics}
\graphicspath{{logos/}{figures/}}

\usepackage{ANA-HIGP-2024-26-PAPER-defs}

\AtlasTitle{Combination of measurements of CP properties of Higgs boson interactions with vector bosons using proton--proton collisions at $\sqrt{s} = 13$~TeV with the ATLAS detector}

\AtlasAbstract{%
A combination of measurements of the CP properties of Higgs boson interactions with electroweak gauge bosons is presented, using 140~fb$^{-1}$ of proton--proton collisions at $\sqrt{s} = 13$~TeV recorded by the ATLAS detector. Results from VBF $H\to\tau\tau/WW^{*}/\gamma\gamma$, inclusive $H\to ZZ^{*}$, and $WH,H\to b\bar{b}$ channels are combined.
No evidence of CP violation is observed, and constrains on the CP-violating operators in the Standard Model Effective Field Theory framework are set in the Warsaw basis.
The results from the combination improve by over 40\% on previous individual limits on $c_{H\tilde{W}}$ and, for the first time, simultaneous constraints on three coefficients $c_{H\tilde{W}}$, $c_{H\tilde{B}}$, and $c_{H\tilde{W}B}$ are set.
These limits are the most stringent constraints to date on the relevant Wilson coefficients in the SMEFT framework with minimum model dependence.
}

\author{The ATLAS Collaboration}

\AtlasRefCode{HIGP-2024-26}

\PreprintIdNumber{CERN-EP-2026-071}

\arXivId{2603.20117}

\AtlasJournalRef{Phys. Rev. Lett. 137 (2026) 021801}
\AtlasDOI{10.1103/54vv-lgc7}

\AtlasCoverEgroupAnalysisTeam{atlas-higp-2024-26-analysis-team@cern.ch}

\hypersetup{pdftitle={ATLAS document},pdfauthor={The ATLAS Collaboration}}

\begin{document}
\maketitle

%

%
%

%
The Standard Model (SM) of particle physics is unable to explain the observed baryon asymmetry of the universe. One of the three Sakharov conditions~\cite{Sakharov:1967dj} required to generate this asymmetry is the violation of charge--parity (CP) symmetry. In the SM, CP violation arises from the complex phase of the quark mixing matrix~\cite{Kobayashi:1973fv,Cabibbo:1963yz}, and possibly the phase of neutrino mixing. However these effects are insufficient to account for the observed baryon asymmetry, and additional beyond-the-SM (BSM) sources of CP violation are needed.
The discovery in 2012~\cite{HIGG-2012-27,CMS-HIG-12-028} of the Higgs boson ($H$) has opened new possibilities for search of new CP-violating interactions. The Higgs boson is predicted by the SM to be a CP-even scalar and its CP properties were extensively studied~\cite{HIGP-2024-09, HIGP-2024-07, HIGG-2020-08, HIGG-2019-13,HIGG-2019-01, HIGG-2018-30, HIGG-2018-14,HIGG-2017-13,CMS-HIG-20-006,CMS-HIG-20-007,CMS-HIG-19-013,CMS-HIG-19-009,CMS-HIG-17-034,CMS-HIG-22-008,Bahl:2022yrs}. Current measurements are consistent with SM expectations,  but experimental results cannot fully exclude the possibility of CP-violating interactions. Measuring the CP structure of Higgs boson couplings to electroweak gauge bosons constitutes a crucial test of the SM since any deviation from the SM prediction would be a clear sign of BSM.
CP-violating interactions between the Higgs boson and electroweak gauge bosons ($HVV$, $V=W^\pm,Z$) can be probed experimentally through vector boson fusion (VBF), associated production ($VH$), and Higgs boson decays into $ZZ^*$ and $WW^*$.
The Standard Model Effective Field Theory (SMEFT) formalism~\cite{SMEFTsim, Buchmuller:1985jz, Brivio:2017vri, Warsaw} extends the SM Lagrangian with dimension-6 operators to incorporate CP-violating components in $HVV$ interactions, parameterized by Wilson coefficients $c_i$ and a new physics scale $\Lambda$.
The resulting cross-section consists of three components: the SM contribution, a linear term in $c_i$ describing CP-odd interference between SM and BSM amplitudes that is suppressed by $1/\Lambda^2$, and a quadratic term in $c_i$ describing CP-even BSM contributions suppressed by $1/\Lambda^4$.
In addition, cross-terms from different operators can be present.
The linear term affects the shapes of CP-odd observables while the quadratic term impacts CP-even observables such as inclusive cross-sections.
The SMEFT cross-section can be decomposed as:

\begin{equation}
O = O^{\mathrm{SM}}
\left(
1
+ \sum_i A_{i}\, \frac{c_i}{\Lambda^2}
+ \sum_i B_{i}\, \frac{c_i^2}{\Lambda^4}
+ \sum_{i<j} C_{i j}\, \frac{c_i c_j}{\Lambda^4}
\right),
\end{equation}

where $O^{\mathrm{SM}}$ is the corresponding SM prediction. The indices $i,j$ run through the BSM operators considered, while $A_{i}$, $B_{i}$, and $C_{i j}$ parameterize the linear, quadratic, and cross-term contributions, respectively, defined relative to the SM expectation.
Constraints on Wilson coefficients of CP-violating operators are derived in two scenarios: using linear-only or both linear and quadratic terms. %
While the linear-only scenario provides a direct CP-violation test via the measurement of asymmetries in CP-odd observables, the linear plus quadratic scenario provides a more complete assessment of the BSM impact and the validity of the SMEFT truncation.
In the Warsaw basis~\cite{Warsaw}, considering only CP-violating operators that conserve lepton and baryon number, $HVV$ interactions are described by three Wilson coefficients: $\cHWtil$, $\cHBtil$, and $\cHWBtil$.

This Letter presents a combination of constraints on CP violation based on measurements performed in VBF \HGam~\cite{HIGG-2020-08},  VBF \Htautau~\cite{HIGP-2024-09}, \HWW~\cite{HIGP-2024-07}, and \HZZ~\cite{HIGG-2018-30}, and a new measurement of associated production in \WHbb, using 140~\ifb proton--proton collision data at \sqs~=~13~\TeV recorded during Run 2 of the LHC with the ATLAS detector from 2015 to 2018. The results include single-parameter fits for individual constraints on $\cHWtil$ while fixing $\cHBtil$ and $\cHWBtil$ to zero, and simultaneous fits for constraining the coefficients $\cHWtil$, $\cHBtil$, and $\cHWBtil$.

ATLAS~\cite{PERF-2007-01,GENR-2019-02} is a multipurpose particle detector with a forward--backward symmetric cylindrical geometry and a near \(4\pi\) coverage in solid angle. A software suite~\cite{SOFT-2022-02} is used in data simulation, in the reconstruction and analysis of real and simulated data, in detector operations, and in the trigger and data acquisition systems of the experiment.
%
%
%
%
%
%


%
In all input measurements the signal is modeled using Monte Carlo (MC) simulations from \POWHEGBOX[v2]~\cite{Hamilton:2013fea,Hamilton:2015nsa,Alioli:2010xd,Nason:2004rx,Frixione:2007vw,Nason:2009ai} interfaced with \PYTHIA[8]~\cite{Sjostrand:2014zea} for describing SM Higgs boson production via gluon--gluon fusion (ggF), VBF and $WH$, with further details provided in the relevant publications~\cite{HIGG-2020-08,HIGP-2024-09,HIGP-2024-07,HIGG-2018-30}.
BSM CP-odd Higgs bosons are generated with \MADGRAPH~\cite{Alwall:2014hca} at leading-order (LO) in QCD, using \textsc{SMEFTsim3.0}~\cite{Brivio:2017btx, SMEFTsim} in the top flavor assumption and the ${m_W,m_Z,G_F}$ input scheme. The scale $\Lambda$ is set to 1~\TeV following convention.

%
%


%
%
Tests of CP-invariance are performed by all analysis with the use of CP-odd observables.  Under CP conjugation, a CP-odd observable $\mathcal{O}$ transforms as $CP(\mathcal{O})=-\mathcal{O} $. If CP-invariance holds, the average over the full phase space vanishes: $\langle \mathcal{O} \rangle = 0$. Any asymmetry in the distribution of $\mathcal{O} $ would then be a sign of CP-violation.  All analysis use a \textit{shape-only} approach, where the normalisation of the signal sample is added as a free floating parameter. In this way, the analyses are not sensitive to overall changes of the total cross-section from CP-even terms, higher-order BSM corrections~\cite{Degrande:2016dqg} can be factored out. Only the differences in the shape of the CP-sensitive variables are considered. %

In the following paragraphs, the main characteristics of each input channel are summarized, and a previously unpublished analysis in the \WHbb channel is described.

In \HGam, \Htautau, and \HZZ channels, the Optimal Observable (\OO) method~\cite{Atwood:1991ka,Davier:1992nw,Diehl:1996wm} is employed, as they have fully reconstructed final state particles. An \OO is defined as $2\text{Re}(\mathcal{M}_\text{SM}^{*}\mathcal{M}_\text{BSM})/|\mathcal{M}_\text{SM}|^2$,
where $\mathcal{M}_\text{SM}$ and $\mathcal{M}_\text{BSM}$ denote SM and BSM matrix elements respectively.
The \OO condenses multidimensional information into a single observable with optimal sensitivity to CP-violating contributions.  %
The \HZZ analysis in Ref.~\cite{HIGG-2018-30} constructs two \OO for the single-parameter fit: a production observable, $OO_{jj}^{\cHWtil}$, used in the VBF signal region with $\mathcal{M} (VBF)$, and a decay observable, $OO_{4l}^{\cHWtil}$, used in a VBF-depleted signal region with $\mathcal{M} (\HZZ)$.
In the simultaneous fit, two decay observables $OO_{4l}^{\cHWtil}$ and $OO_{4l}^{\cHBtil}$ related to the two coefficients \cHWtil, \cHBtil are regarded as observables in the inclusive \HZZ signal region, providing optimal sensitivity to the three Wilson coefficients.
In addition to what was published in Ref.~\cite{HIGG-2018-30}, a linear-only interpretation is also performed here.
For the single-parameter linear-only fit of \cHWtil, the \OO distributions are rebinned to avoid bins in which the negative interference contribution exceeds the total background yield, reducing the expected sensitivity by up to 8\%. For \Htautau, an additional simultaneous parameterization of $\cHWtil$, $\cHBtil$, and $\cHWBtil$ is performed compared with Ref.~\cite{HIGP-2024-09}, including interference terms between the different CP-violating amplitudes. %


The \HWW and \WHbb analyses rely on modifications to the simplified template cross-section (STXS) formalism~\cite{STXS-LesHouches,STXS-Stage1.1,deFlorian:2016spz} with dedicated angular observables sensitive to the CP structure of the $HVV$ coupling. In \HWW the CP-sensitive variable is the pseudorapidity-ordered azimuthal angular difference between the two leading jets, \DPhijj, sensitive only to the structure of the interaction at the production vertex. %
In contrast to the result in Ref.~\cite{HIGP-2024-07}, a single-parameter fit of \cHWtil is performed, with the remaining coefficients set to zero.
In addition, the quadratic parametrization of \cHWtil in the single-parameter fit, as well as the quadratic and cross terms in the simultaneous fit, are included only in this combination.

The \WHbb analysis is based on the measurement in Ref.~\cite{HIGG-2020-20}, where events are selected to reconstruct a leptonic $W$-boson decay and a $H$ decay into $b$-quarks. To reconstruct the Higgs boson decay products, small-radius jets are used for $\pTW<400~\GeV$ and large-radius jets for $\pTW>400~\GeV$, where \pTW~is the transverse momentum of the $W$ boson. The $W$-boson four-vector is reconstructed as the vector sum of the charged lepton and the neutrino, where the transverse component of the neutrino is identified with the \MET and the longitudinal component is obtained by applying a $W$-mass constraint to the lepton-neutrino system.
MC simulated samples model most of the SM background processes, and dedicated control regions are used to constrain the main backgrounds ($W$+jets and top quark production).
Compared with the STXS categories used in Ref.~\cite{HIGG-2020-20}, the regions are further split into two bins of an angular variable, \qCDP, where $Q_\ell$ is the lepton charge and $\cos \delta^+$ is an angular observable defined as $\frac{\mathbf{p}_\ell^{(W)} \cdot \left( \mathbf{p}_H \times \mathbf{p}_W \right)}{|\mathbf{p}_\ell^{(W)}| \cdot |\mathbf{p}_H \times \mathbf{p}_W| }$.
Here, $\mathbf{p}_{\ell}^{(W)}$ is the three-momentum of the charged lepton in the $W$-boson rest frame, and $\mathbf{p}_W$ and $\mathbf{p}_H$ are the three-momenta of the $W$ and Higgs bosons, respectively, in the laboratory frame~\cite{GodboleHCP,sally_cpv_wh}.
The resulting cross-section measurements are shown in Figure~\ref{fig:stxs_whbb} in bins of the particle-level transverse momentum of the $W$ boson, $\pTWt$, and \qCDP.  Compared with the result in Ref.~\cite{HIGG-2020-20}, only $WH$ production, with $W\to \ell \nu\ (\ell = e,\mu)$, is considered and $ZH$ is treated as a background. Events with hadronically decaying $\tau$-leptons are vetoed,
and only one \pTW~bin above 400~\GeV is considered to reduce the impact of statistical uncertainties.

The EFT interpretation is performed by only considering the \qCDP shape effect from \cHWtil.
One signal normalization factor is applied per \pTW~bins and  profiled in fitting, while the normalization factors in negative and positive \qCDP bins are parameterized with \cHWtil.

\begin{figure}[h!]
\centering
\includegraphics[width=0.6\linewidth]{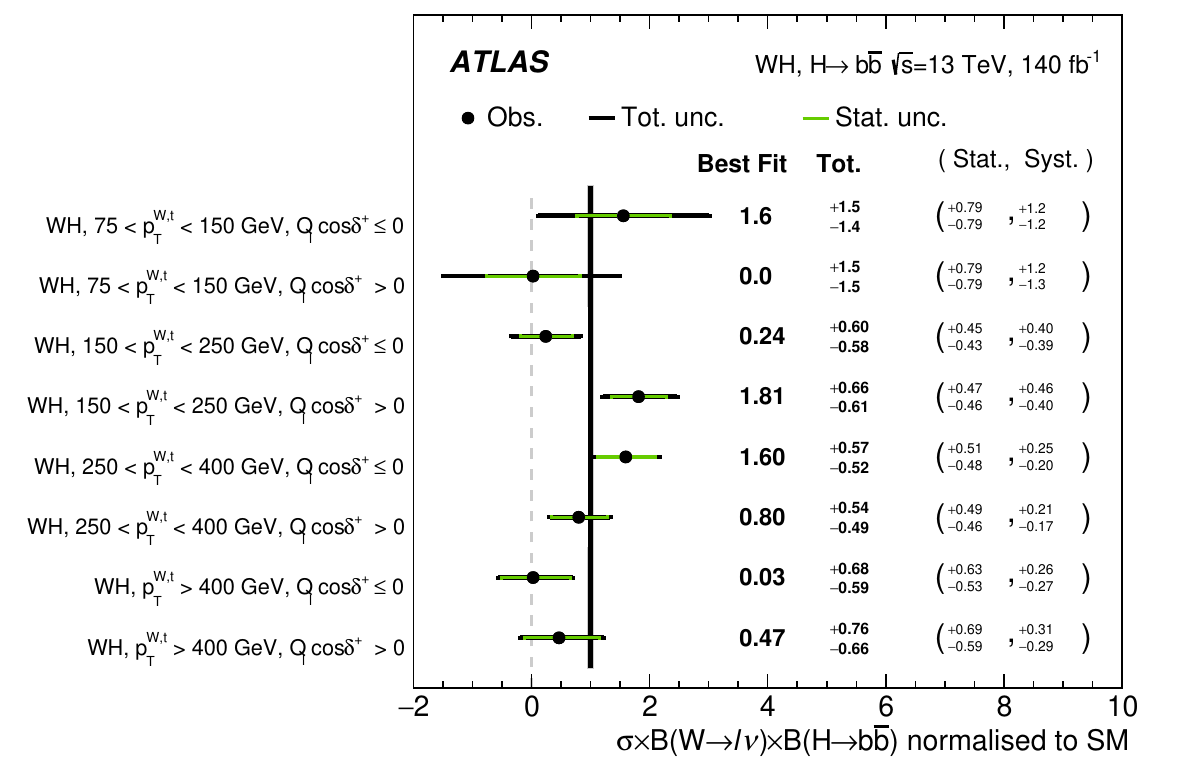}
\caption{Measured $WH$ production cross-sections times the $W\rightarrow \ell \nu$ and the $H \rightarrow b\bar{b}$ branching ratios normalized to their SM predictions. The best fit values and corresponding uncertainties (total as well as its statistical and systematic components) are shown for all regions. Theoretical uncertainties on the predictions are not shown as they are too small to be visible.}%
\label{fig:stxs_whbb}
\end{figure}



%
%
%
%

%
%
%
%

Combination results are presented using single-parameter fits for individual constraints on $\cHWtil$ (with \mbox{$\cHBtil = \cHWBtil=0$}) and simultaneous fits for constraining $\cHWtil$, $\cHBtil$ and $\cHWBtil$. The VBF production process is mainly sensitive to \cHWtil, with subleading contributions from \cHWBtil and \cHBtil, while $WH$ production depends only on \cHWtil. %
The simultaneous fits combine all channels except for \HGam, which has very limited sensitivity to the additional operators and has a negligible impact on the simultaneous fit results.
The normalization factors are treated as free parameters for all production modes for all channels.
The exceptions are the normalization of $ZH$ production in the \WHbb channel, and the ggF production mode in the \WHbb and VBF \Htautau channels, which are constrained to their respective SM prediction, since these are minor contributions.
The relative ratio between VBF and ggF production in the \HZZ channel is also constrained to the SM prediction, since this channel only uses the decay information.
All free normalization factors are uncorrelated across channels.
Event overlap between channels is negligible due to their distinct final states and selection criteria.
Signal injection tests with non-zero Wilson coefficients were performed to varify the extraction of Wilson coefficients is unbiased in this shape-only approach.

The combined likelihood is constructed as the product of the individual analyses likelihoods, incorporating correlations of systematic uncertainties where applicable. The treatment of systematic uncertainties follows that of the individual analyses~\cite{HIGG-2020-08,HIGP-2024-09,HIGP-2024-07,HIGG-2018-30}, with experimental uncertainties related to luminosity, object reconstruction and calibration considered fully correlated across all channels. Theoretical uncertainties affecting signal modeling are correlated among analyses targeting the same production mode, while background modeling uncertainties are treated as uncorrelated due to their analysis-specific nature. The impact of systematic uncertainties in the presented results is small compared to the statistical uncertainties. In each fit, the confidence intervals for the Wilson coefficients are determined using a profile likelihood ratio test statistic as in Ref.~\cite{HIGG-2018-30}.

No evidence of CP-violation is observed, and constraints on the Wilson coefficients are set. The observed and expected negative log-likelihood ratio ($2\Delta$NLL) scans as a function of the \cHWtil~are shown in Figure~\ref{fig:1d} for the single-parameter fit (\cHBtil~=~\cHWBtil~=~0), separately for linear-only and linear plus quadratic scenarios. Additionally, the scans obtained in simultaneous fits with all coefficients floating are also shown. The sensitivity to \cHWtil~in this case is impacted by the negative correlation with the remaining coefficients. In the single-parameter fit, observed (expected) 95\% CL intervals of \intOBSlin~(\intEXPlin) and \intOBSlinquad~(\intEXPlinquad) at $\Lambda = 1~\TeV$ are obtained for the linear-only and linear plus quadratic scenarios, respectively.  The limits obtained on \cHWtil in the single-parameter fit show an improvement of over 40\% on the previous best individual channel results~\cite{HIGP-2024-09}. The expected and observed limits are primarily driven by \Htautau, followed by the \WHbb channel.

\begin{figure}[h!]
\centering
\subfloat[]{\includegraphics[width=0.4\linewidth]{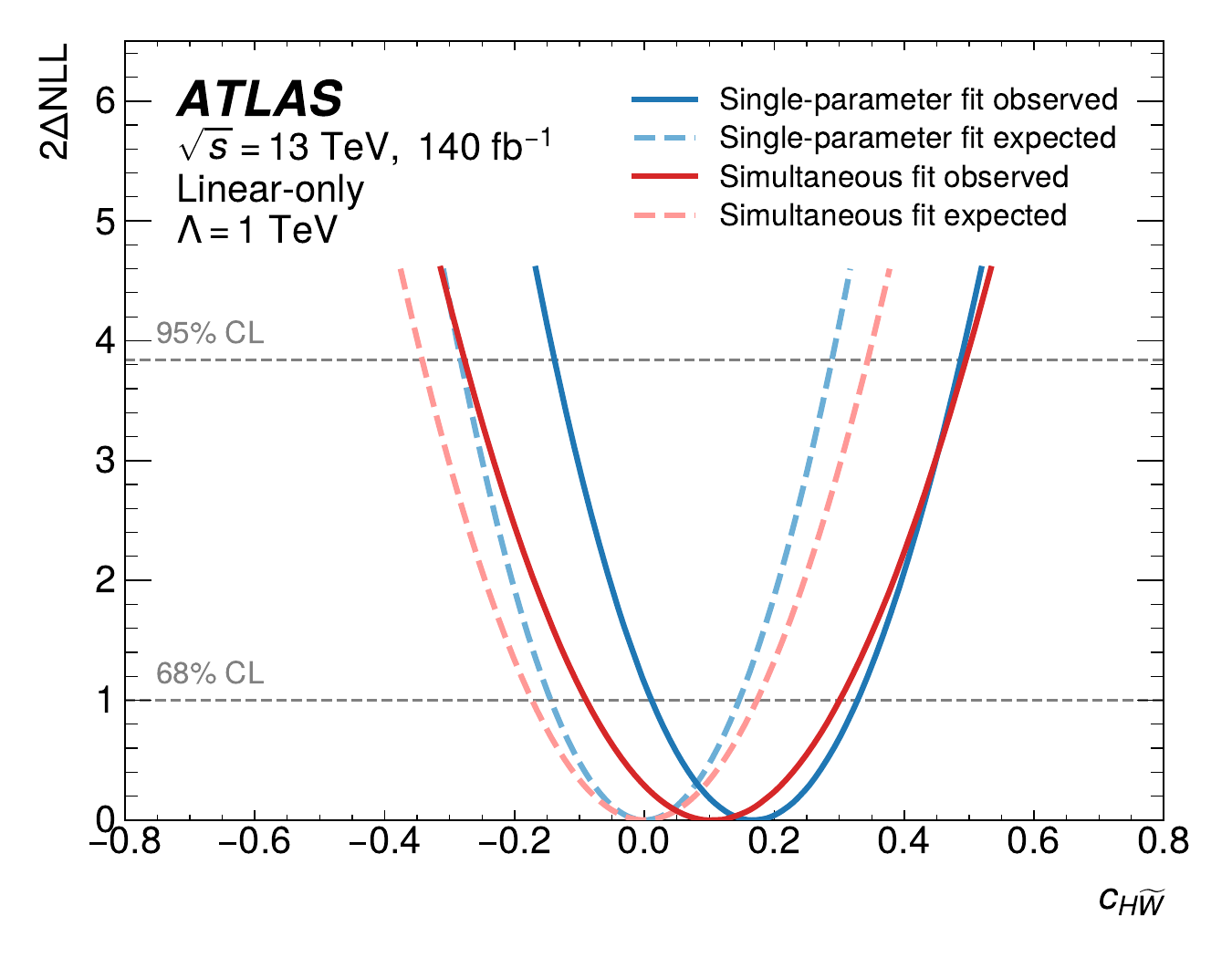}}
\subfloat[]{\includegraphics[width=0.4\linewidth]{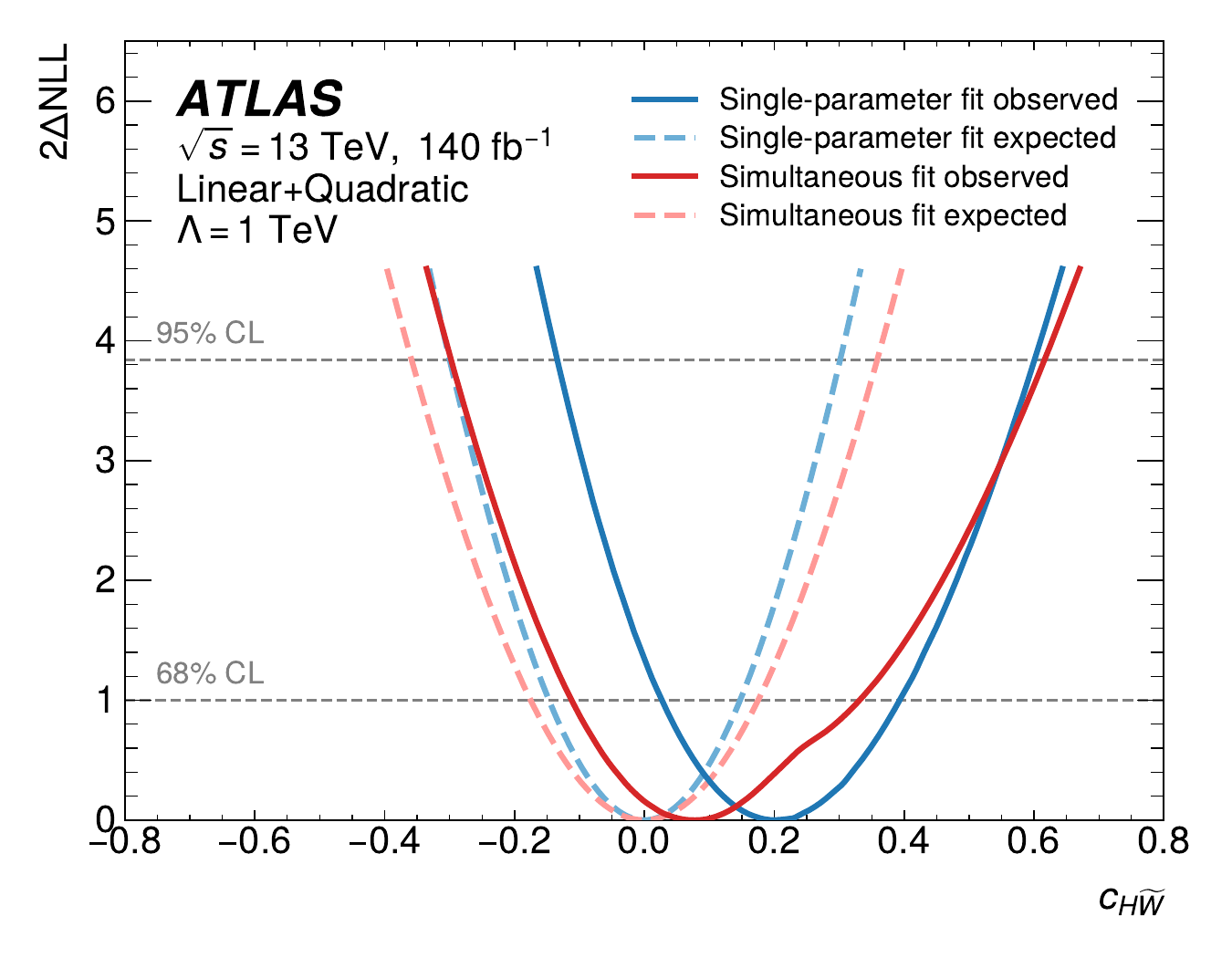}}
\caption{Negative profile log-likelihood ratio scans are shown as a function of \cHWtil, with single-parameter fits (\cHBtil~=~\cHWBtil~=~0) and simultaneous fits (all three coefficients floating), for (a) linear-only and (b) linear plus quadratic terms. Solid lines indicate the observed scans, while dashed lines show the expected results. The dashed horizontal lines show the thresholds defining the 68\% and 95\% confidence intervals, assuming the asymptotic approximation.}
\label{fig:1d}
\end{figure}


Results of the simultaneous fits are shown in Figure~\ref{fig:3d} in terms of two-dimensional contours for each pair of Wilson coefficients, profiling over the third coefficient. There is a large correlation between the \cHBtil and \cHWBtil coefficients caused by similar effects to the CP-odd observables in the \HZZ, \HWW and \Htautau channels. On the other hand, \cHWtil is positively correlated with \cHWBtil and negatively correlated with \cHBtil in the \HZZ channel, negatively correlated with both coefficients in \HWW and \Htautau, and independent in the \WHbb channel. There are parameter-space regions where the linear-only terms lead to negative cross-sections; however, these are outside the observed 95\% CL contours and outside of EFT validity regions.
The limits obtained in the simultaneous fit are less stringent than those obtained in the single-parameter fit, due to the negative correlation between \cHWtil and the remaining coefficients.
Figure~\ref{fig:3POI_best_fit} summarizes the observed results obtained for each Wilson coefficient when profiling the other two for both linear-only and linear plus quadratic scenarios.

\begin{figure}[h!]
\centering
\subfloat[]{ \includegraphics[width=0.31\linewidth]{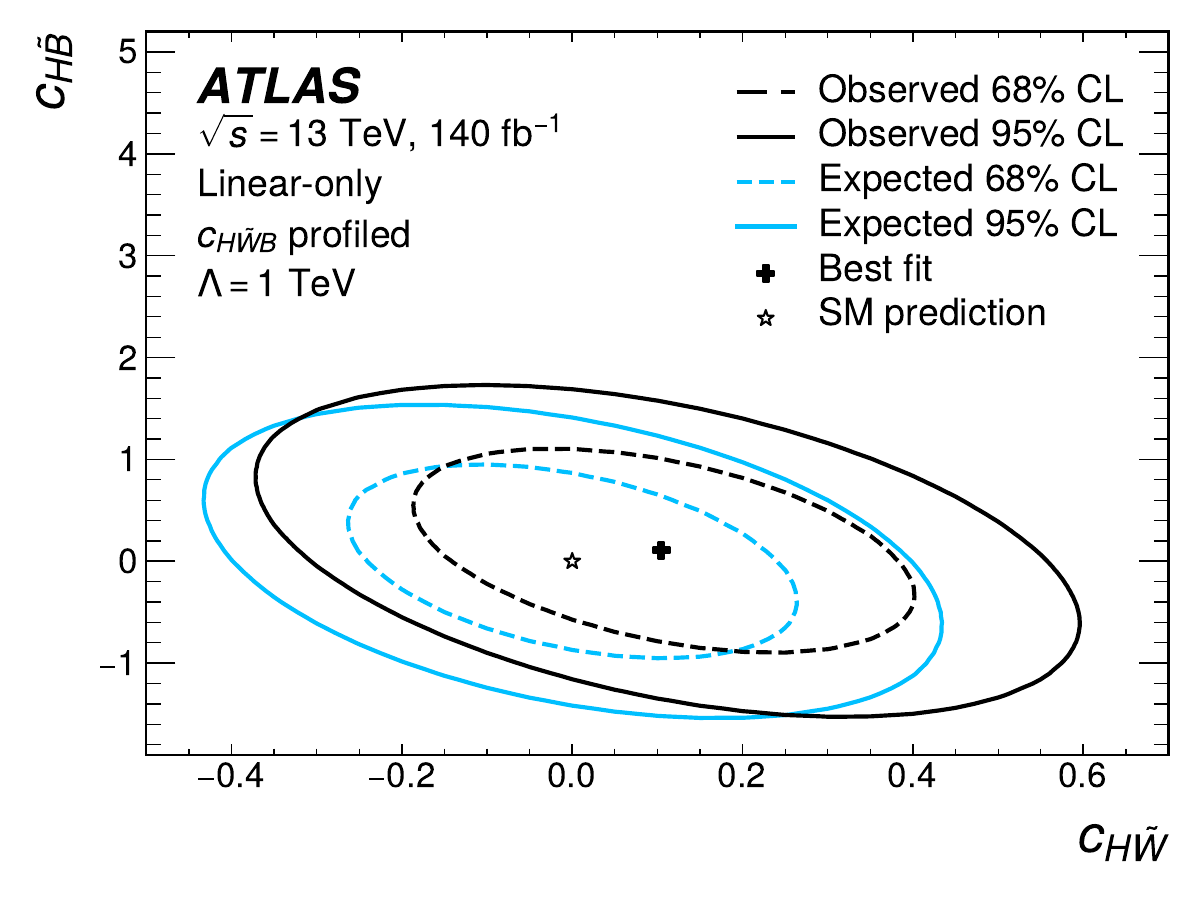}}
\subfloat[]{ \includegraphics[width=0.31\linewidth]{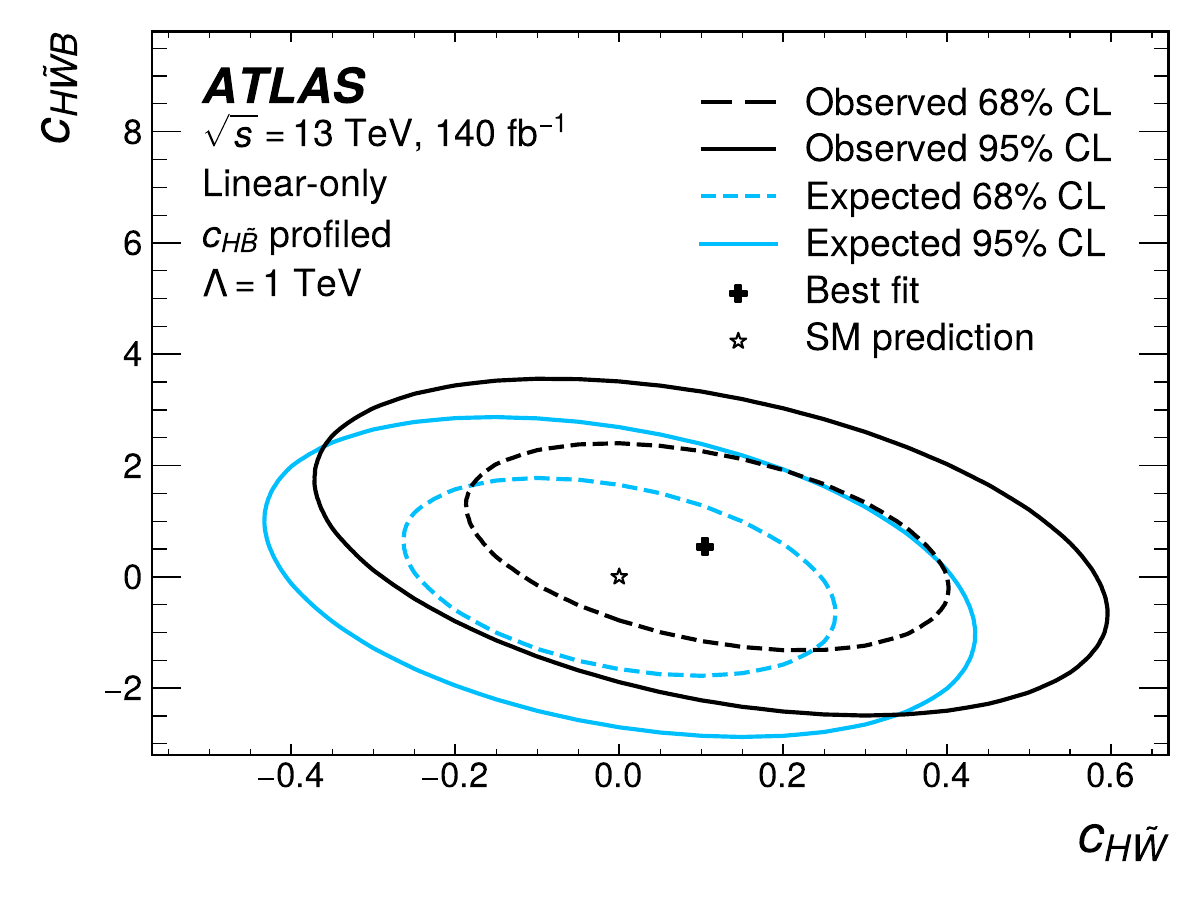}}
\subfloat[]{ \includegraphics[width=0.31\linewidth]{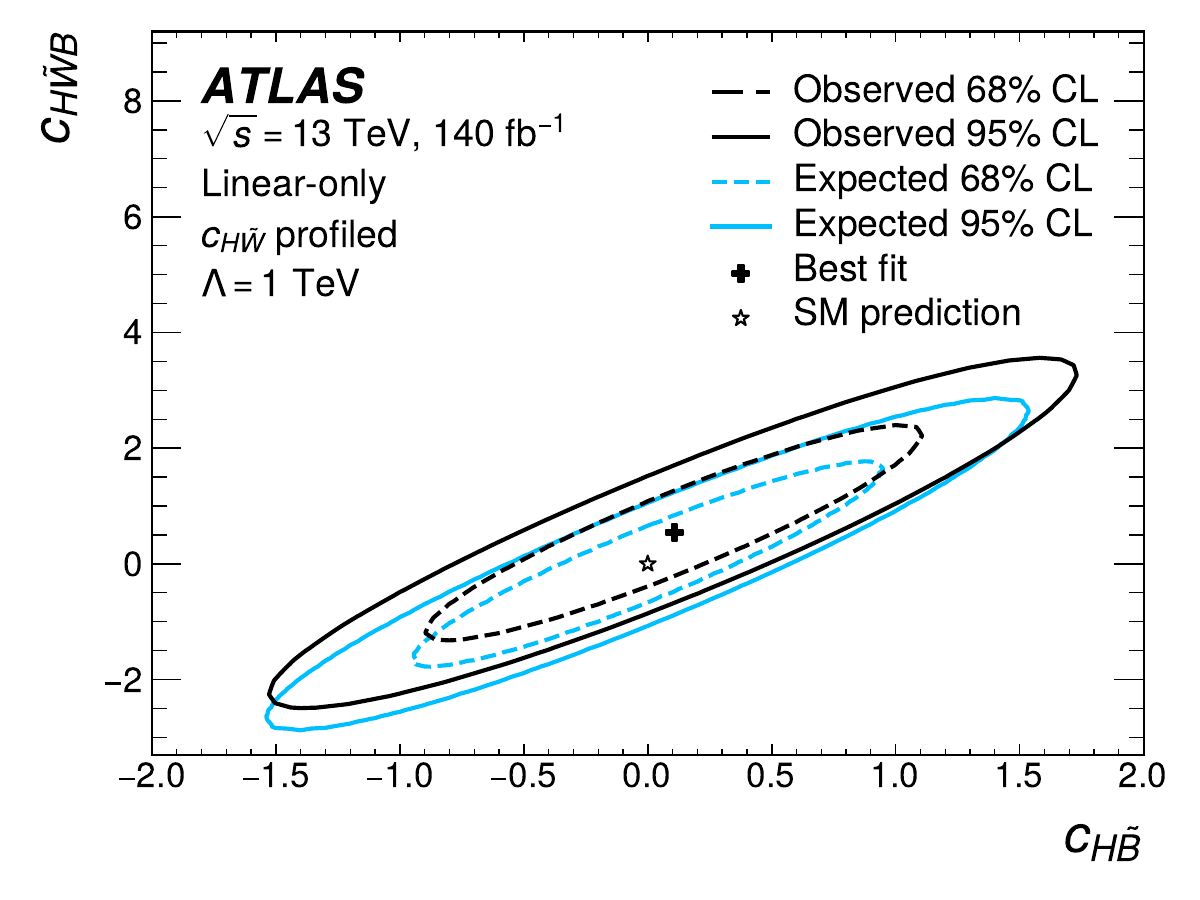}}\\
\subfloat[]{\includegraphics[width=0.31\linewidth]{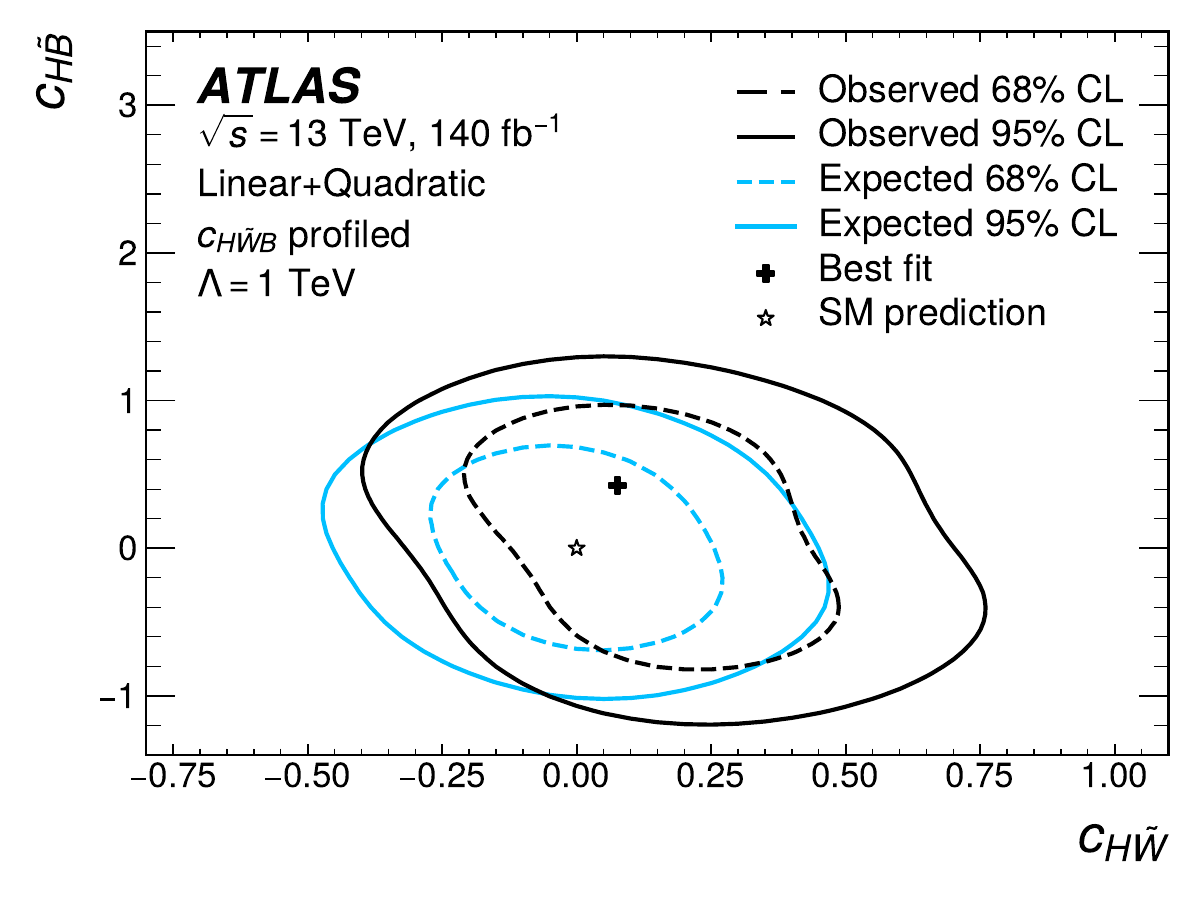}}
\subfloat[]{  \includegraphics[width=0.31\linewidth]{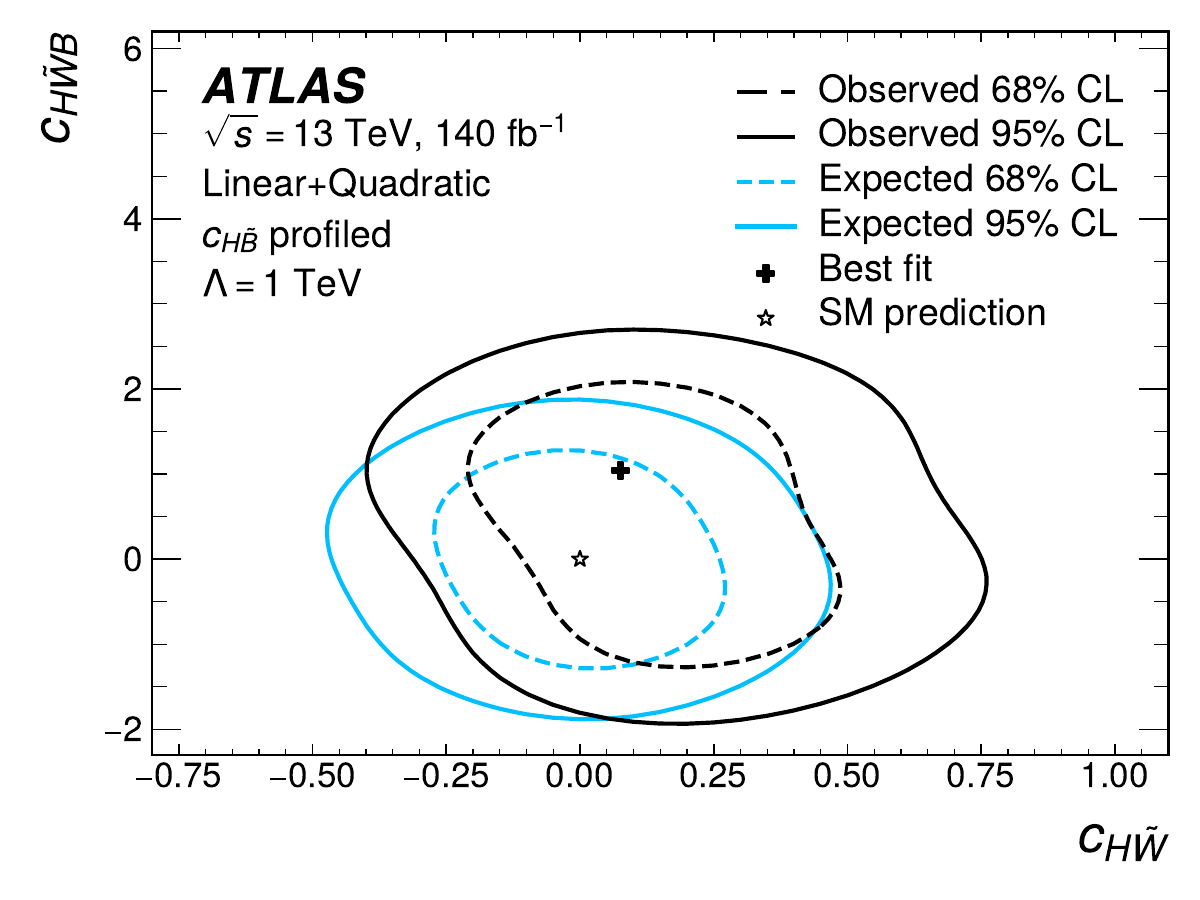}}
\subfloat[]{  \includegraphics[width=0.31\linewidth]{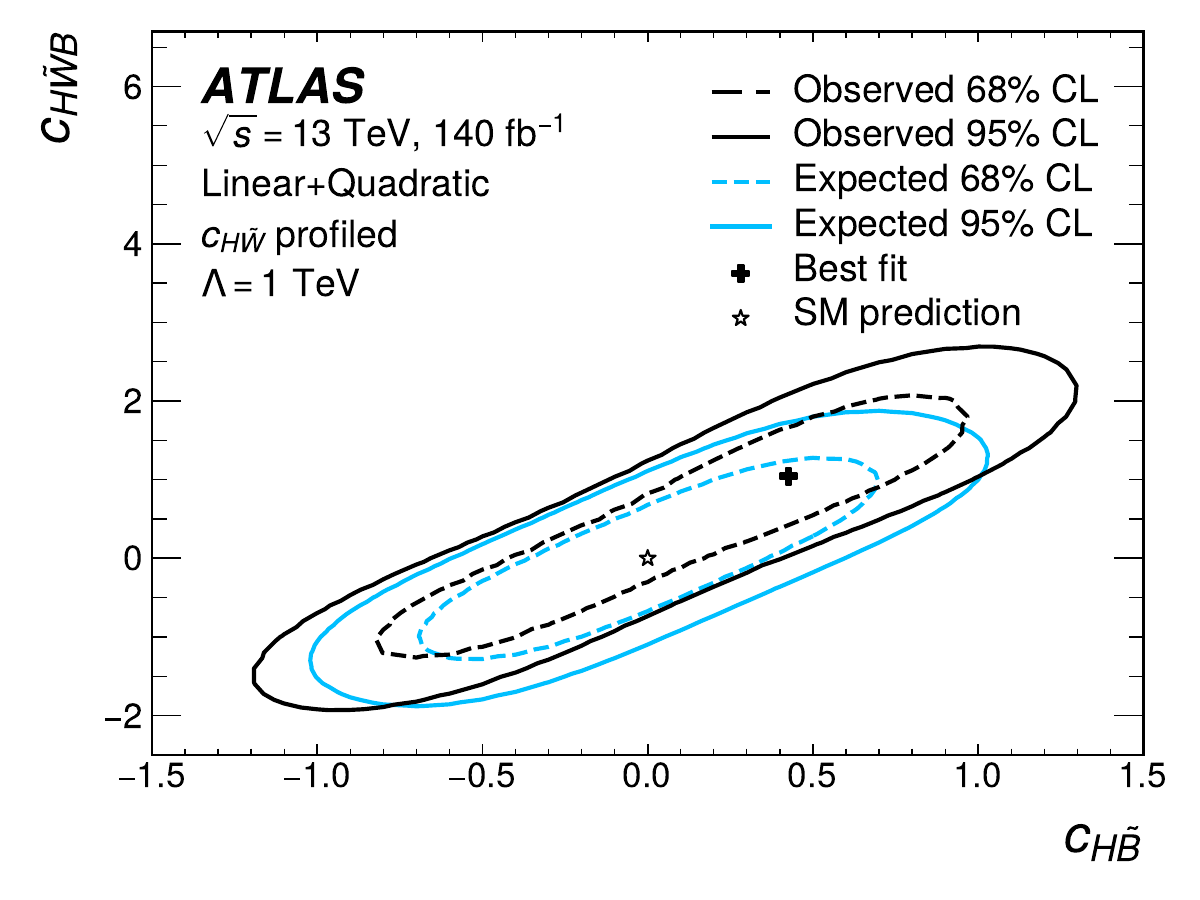}}
\caption{The observed 68\% (dashed line) and 95\% (solid line) CL two-dimensional contours are shown for all three pairings of the Warsaw basis: (a, d)  \cHWtil versus \cHBtil, (b, e) \cHWtil versus \cHWBtil, and (c, f) \cHBtil versus \cHWBtil a simultaneous fit, for (a, b, c) linear-only and (d, e, f) linear plus quadratic terms interpretations. Expected contour lines are also shown. All couplings scale as $1/\Lambda^2$ with $\Lambda = 1$~\TeV.}
\label{fig:3d}
\end{figure}

\begin{figure}[h!]
\centering
\includegraphics[width=0.5\linewidth]{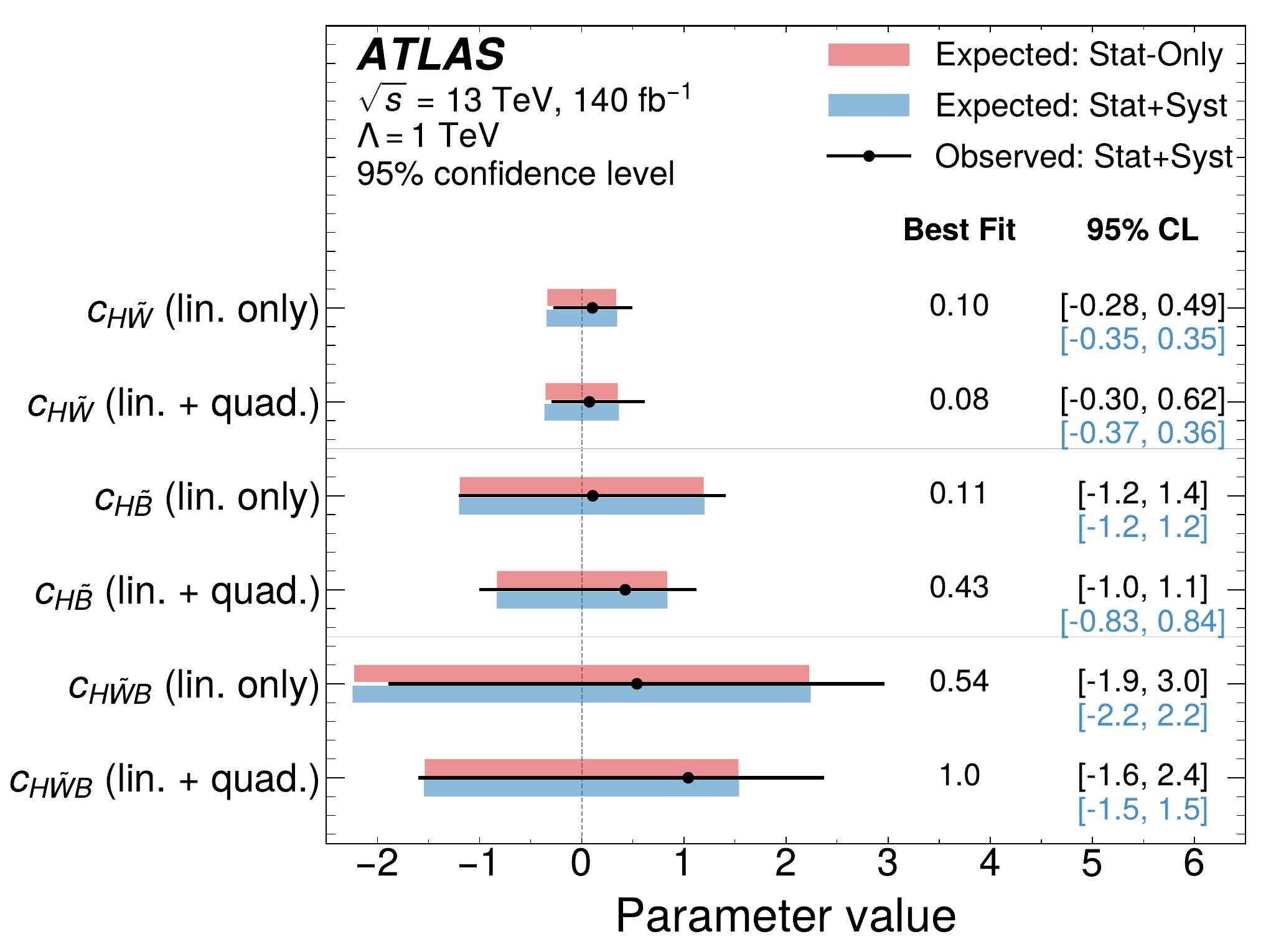}
\caption{Best-fit values and 95\% CL intervals for all three Wilson coefficients obtained in a simultaneous fit to all coefficients. In both cases, linear and linear plus quadratic scenarios are shown. Expected limits using the full model and statistical-only are also presented.}
\label{fig:3POI_best_fit}
\end{figure}


In conclusion, constraints of Wilson coefficients of CP-violating operators in the $HVV$ interaction within the SMEFT Warsaw basis are presented. %
The analysis combines five ATLAS Run~2 measurements including VBF and $WH$ production modes, and $H\to VV$ decay channels: \Htautau, \HWW, \HGam, \HZZ, and \WHbb. Shape-based measurements of CP-odd sensitive observables are used to constrain CP-violating effects. A single-parameter fit was performed constraining the $\cHWtil$ Wilson coefficient while the remaining coefficients are set to zero. No significant deviations from the Standard Model are observed. The single-parameter fit yields an observed (expected) 95\% CL interval for \cHWtil of \intOBSlin~(\intEXPlin) at $\Lambda=1~\TeV$ for the linear-only and \intOBSlinquad~(\intEXPlinquad) for the linear plus quadratic interpretation.  The compatibility between the limits in both scenarios is an indication of the validity of the EFT truncation to dimension-six operators.
The limits presented on \cHWtil from the single-parameter fit improve upon previous best individual channel results from  \Htautau~\cite{HIGP-2024-09} by over 40\%.
The combination further enabled, for the first time, a simultaneous fit constraining the coefficients of three CP-violating operators ($\cHWtil$, $\cHBtil$, $\cHWBtil$).
These combined results deliver the most stringent constraints to date on Wilson coefficients of CP-violating operators in the SMEFT framework.


\clearpage
\section*{Acknowledgements}

%

%
%

%
%

We thank CERN for the very successful operation of the LHC and its injectors, as well as the support staff at
CERN and at our institutions worldwide without whom ATLAS could not be operated efficiently.

The crucial computing support from all WLCG partners is acknowledged gratefully, in particular from CERN, the ATLAS Tier-1 facilities at TRIUMF/SFU (Canada), NDGF (Denmark, Norway, Sweden), CC-IN2P3 (France), KIT/GridKA (Germany), INFN-CNAF (Italy), NL-T1 (Netherlands), PIC (Spain), RAL (UK) and BNL (USA), the Tier-2 facilities worldwide and large non-WLCG resource providers. Major contributors of computing resources are listed in Ref.~\cite{ATL-SOFT-PUB-2026-001}.

We gratefully acknowledge the support of ANPCyT, Argentina; YerPhI, Armenia; ARC, Australia; BMWFW and FWF, Austria; ANAS, Azerbaijan; CNPq and FAPESP, Brazil; NSERC, NRC and CFI, Canada; CERN; ANID, Chile; CAS, MOST and NSFC, China; Minciencias, Colombia; MEYS CR, Czech Republic; DNRF and DNSRC, Denmark; IN2P3-CNRS and CEA-DRF/IRFU, France; SRNSFG, Georgia; BMFTR, HGF and MPG, Germany; GSRI, Greece; RGC and Hong Kong SAR, China; ICHEP and Academy of Sciences and Humanities, Israel; INFN, Italy; MEXT and JSPS, Japan; CNRST, Morocco; NWO, Netherlands; RCN, Norway; MNiSW, Poland; FCT, Portugal; MNE/IFA, Romania; MSTDI, Serbia; MSSR, Slovakia; ARIS and MVZI, Slovenia; DSI/NRF, South Africa; MICIU/AEI, Spain; SRC and Wallenberg Foundation, Sweden; SERI, SNSF and Cantons of Bern and Geneva, Switzerland; NSTC, Taipei; TENMAK, T\"urkiye; STFC/UKRI, United Kingdom; DOE and NSF, United States of America.

Individual groups and members have received support from BCKDF, CANARIE, CRC and DRAC, Canada; CERN-CZ, FORTE and PRIMUS, Czech Republic; COST, ERC, ERDF, Horizon 2020 and Marie Sk{\l}odowska-Curie Actions, European Union; Investissements d'Avenir Labex, Investissements d'Avenir Idex and ANR, France; DFG and AvH Foundation, Germany; Herakleitos, Thales and Aristeia programmes co-financed by EU-ESF and the Greek NSRF, Greece; BSF-NSF and MINERVA, Israel; NCN and NAWA, Poland; La Caixa Banking Foundation, CERCA and AGAUR programs from Generalitat de Catalunya and PROMETEO and GenT Programmes Generalitat Valenciana, Spain; G\"{o}ran Gustafssons Stiftelse, Sweden; The Royal Society and Leverhulme Trust, United Kingdom; Eric and Wendy Schmidt Fund for Strategic Innovation, United States of America.

In addition, individual members wish to acknowledge support from Chile: Agencia Nacional de Investigaci\'on y Desarrollo (ANID FONDECYT reg. 1230987, FONDECYT 1230812, FONDECYT 1240864, Fondecyt 3240661, Fondecyt Regular 1240721); China: Chinese Ministry of Science and Technology (MOST-2023YFA1605700, MOST-2023YFA1609300), National Natural Science Foundation of China (NSFC 12275265, NSFC-W2543005); Czech Republic: Czech Science Foundation (GACR - 24-11373S), Ministry of Education Youth and Sports (ERC-CZ-LL2327, FORTE CZ.02.01.01/00/22\_008/0004632), PRIMUS Research Programme (PRIMUS/21/SCI/017); EU: H2020 European Research Council (ERC - 101002463); European Union: European Research Council (BARD No. 101116429, ERC - 948254, ERC 101089007), European Regional Development Fund (HE COFUND GA No.101081355, ERDF), Marie Sklodowska-Curie Actions (GAP-101168829); France: Agence Nationale de la Recherche (ANR-21-CE31-0013, ANR-22-EDIR-0002, ANR-24-CE31-0504-01); Germany: Deutsche Forschungsgemeinschaft (DFG - 469666862); China: Research Grants Council (GRF); Italy: Istituto Nazionale di Fisica Nucleare (LHC-MIUR - 28003/2025), Ministero dell'Università e della Ricerca (NextGenEU 153D23001490006 M4C2.1.1, NextGenEU I53D23000820006 M4C2.1.1, NextGenEU I53D23001490006 M4C2.1.1, SOE2024\_0000023); Japan: Japan Society for the Promotion of Science (JSPS KAKENHI  JP25H0063, JSPS KAKENHI JP22H01227, JSPS KAKENHI JP22H04944, JSPS KAKENHI JP22KK0227, JSPS KAKENHI JP24K23939, JSPS KAKENHI JP24KK0251, JSPS KAKENHI JP25H00650, JSPS KAKENHI JP25H01291, JSPS KAKENHI JP25K01011, JSPS KAKENHI JP25K01023); Poland: Polish National Science Centre (NCN 2021/42/E/ST2/00350, NCN OPUS 2023/51/B/ST2/02507, NCN OPUS nr 2022/47/B/ST2/03059, NCN UMO-2019/34/E/ST2/00393, UMO-2022/47/O/ST2/00148, UMO-2023/49/B/ST2/04085, UMO-2023/51/B/ST2/00920, UMO-2024/53/N/ST2/00869); Spain: Agència de Gestió d'Ajuts Universitaris i de Recerca. (AGAUR - 2023 BP 00141), Ministry of Science and Innovation (RYC2019-028510-I, RYC2020-030254-I, RYC2021-031273-I, RYC2022-038164-I), Ministerio de Ciencia, Innovación y Universidades/Agencia Estatal de Investigaci\'on (EU NextGenerationEU (PRTR-C17.I1), PID2022-142604OB-C22); Sweden: Carl Trygger Foundation (Carl Trygger Foundation CTS 22:2312), Swedish Research Council (Swedish Research Council 2023-04654, VR 2021-03651, VR 2022-03845, VR 2022-04683, VR 2023-03403, VR 2024-05451, VR 2025-05940), Knut and Alice Wallenberg Foundation (KAW 2023.0366); Switzerland: Swiss National Science Foundation (SNSF - PCEFP2\_194658); United Kingdom: The Binks Trust, Royal Society (NIF-R1-231091); United States of America: U.S. Department of Energy (ECA DE-AC02-76SF00515), John Templeton Foundation (John Templeton Foundation 63206), Neubauer Family Foundation.

%
%


\clearpage
\section*{End Matter}

%
%
%

\paragraph{Individual channel contributions to \cHWtil constraint:} The individual negative log-likelihood ratio ($2\Delta\NLL$) scans are shown in Figure~\ref{fig:1d_individual} for linear-only and linear plus quadratic interpretations, and the corresponding best fit values and 95\% CL intervals are summarized in Figure~\ref{fig:1d_and_3d_1poi_combined}, along with the combination results. The introduction of the quadratic term results in a loss of sensitivity at higher Wilson coefficient values, which is visible for \WHbb and to a lesser extent \HWW, but is true for all channels. For \WHbb in particular, this results in undefined confidence intervals.

To understand the non-parabolic behavior of the \NLL scans, it is helpful to consider the particular cases of \WHbb and \HWW. In these two channels, the impact of the CP-odd operators on the corresponding differential cross-section measurements is parameterized via their impact on their signal strengths $\mu_p$ (relative to the SM prediction), defined as:
\begin{equation}
\mu_p\to \mu_p \times \large(1+\sum_{j=1}^{M}A_j^{p}\times c_j + B_j^{p}\times c_j^2 \large),
\end{equation}
where $p$ are the STXS particle-level categories, $A_{j}^{p}$ and $B_{j}^{p}$ parameterize the linear and quadratic contributions in STXS region $p$, and $j$ runs through the number $M$ of CP-odd operators considered. The relative asymmetry across bins of an angular observable can be formulated as $(\mu_{q^+}-\mu_{q^-})/(\mu_{q^+} +\mu_{q^-})$, where $q^\pm$ stands for STXS regions in the same CP-even bin (e.g., \pTW) and symmetric angular observable bins.

The observed effect in the linear plus quadratic \NLL scans is related to how this relative asymmetry depends on the Wilson coefficients. In the linear-only interpretation, this asymmetry grows linearly with \cHWtil (Eq.~\ref{eq:Alinear}); however, in the linear plus quadratic scenario this asymmetry grows approximately linearly only for small values of \cHWtil and is then much reduced (and even inverted) as the absolute value of the coefficient grows (Eq.~\ref{eq:Alinquad}), resulting in the observed \NLL behavior.

\begin{equation}
\text{Asym}^{\text{linear}}_{p,j}=-A_j^{p} \times c_j
\label{eq:Alinear}
\end{equation}

\begin{equation}
\text{Asym}^{\text{linear+quadratic}}_{p,j}=\frac{-A_j^{p} \times c_j}{1+B_j^{p} \times c_j^2}
\label{eq:Alinquad}
\end{equation}

\begin{figure}[h!]
\centering
\subfloat[]{\includegraphics[width=0.4\linewidth]{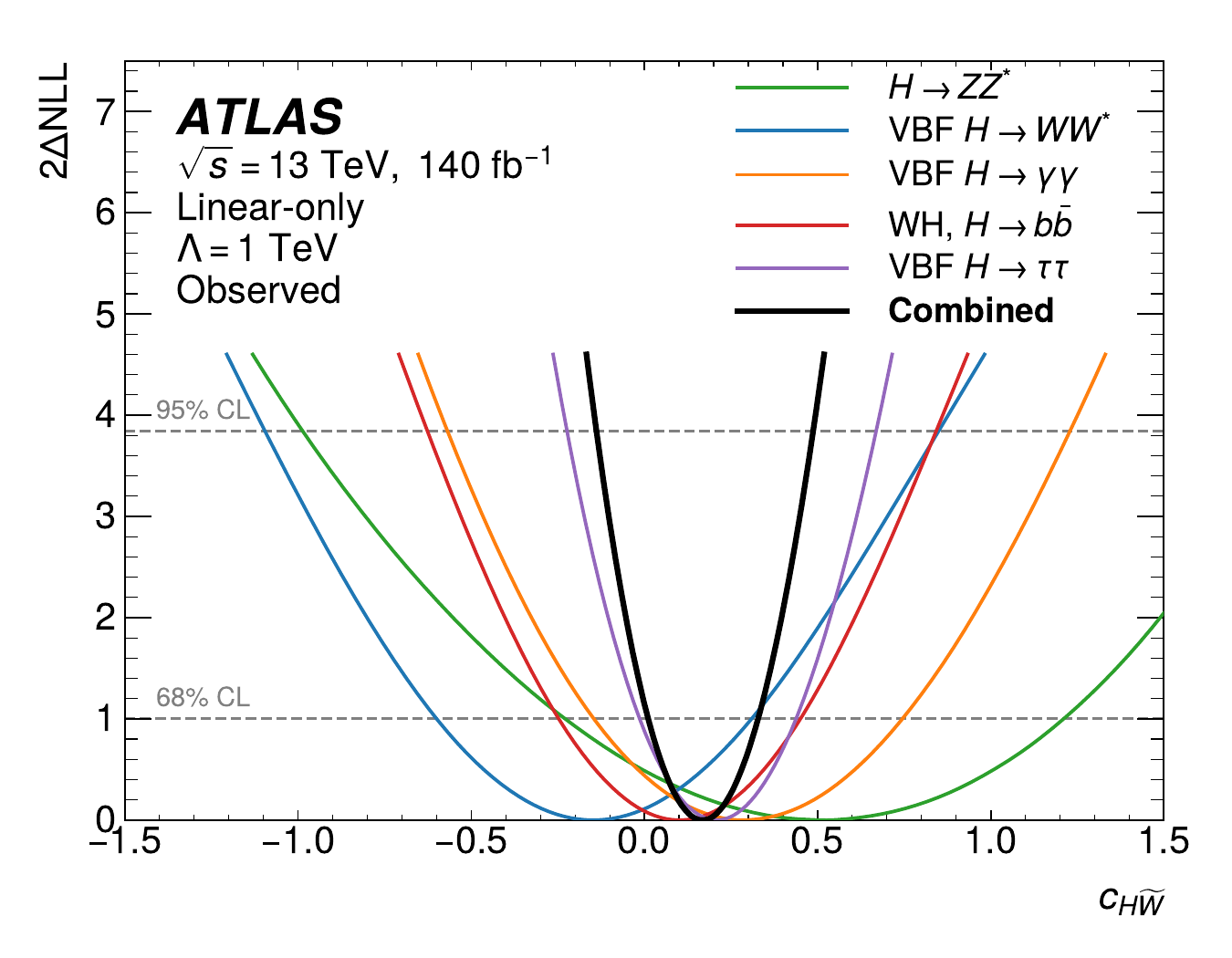}}
\subfloat[]{\includegraphics[width=0.4\linewidth]{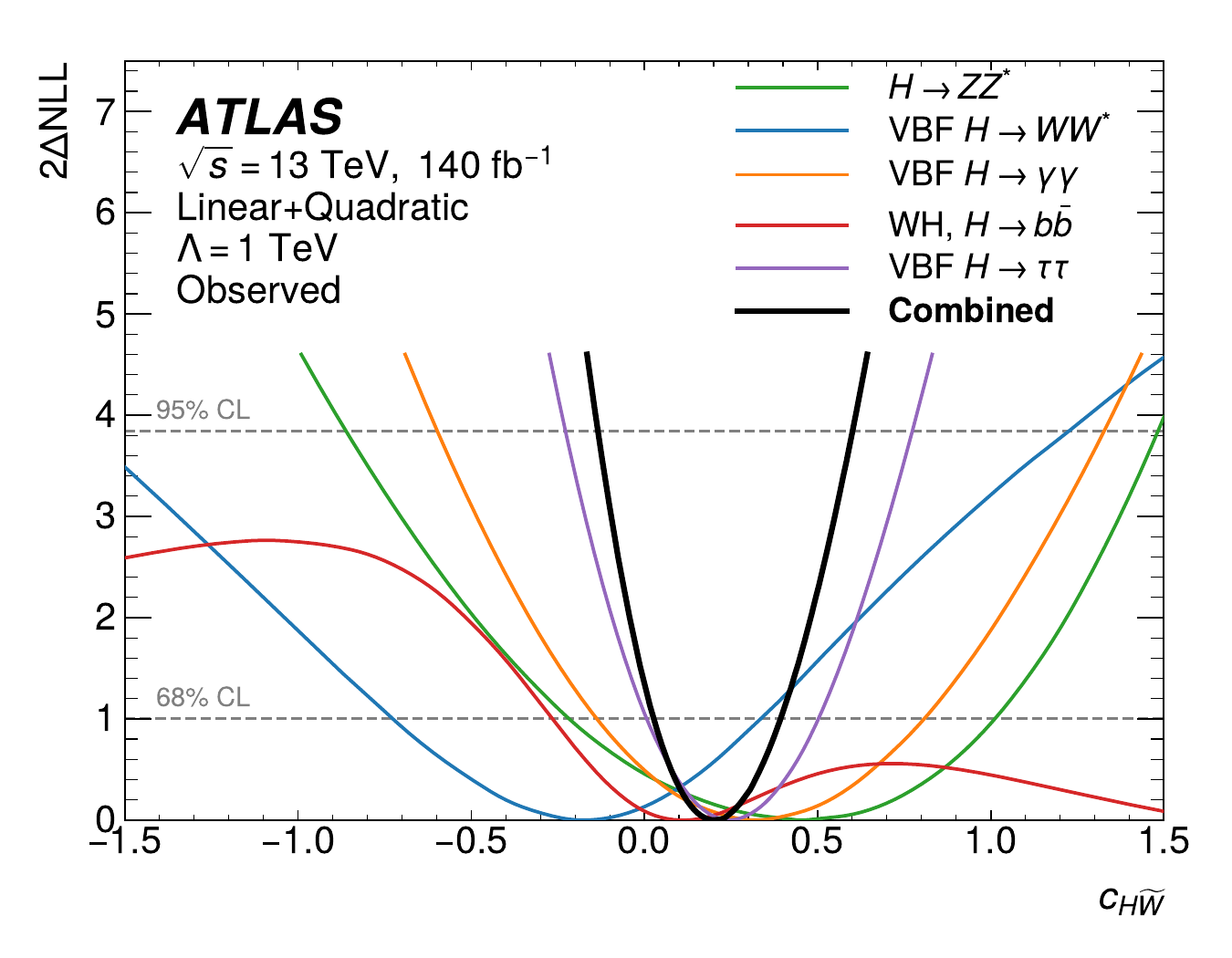}}
\caption{Observed negative log-likelihood ratio ($2\Delta\NLL$) scans as a function of \cHWtil~considering the (a) linear-only and the (b) linear plus quadratic terms for the individual channels (in different colors) and their combination. The dashed horizontal lines show the values of $\Delta\text{NLL}$ thresholds defining the 68\% and 95\% confidence intervals, assuming the asymptotic approximation.}
\label{fig:1d_individual}
\end{figure}

\begin{figure}[h!]
\centering
\subfloat[]{\includegraphics[width=0.46\linewidth]{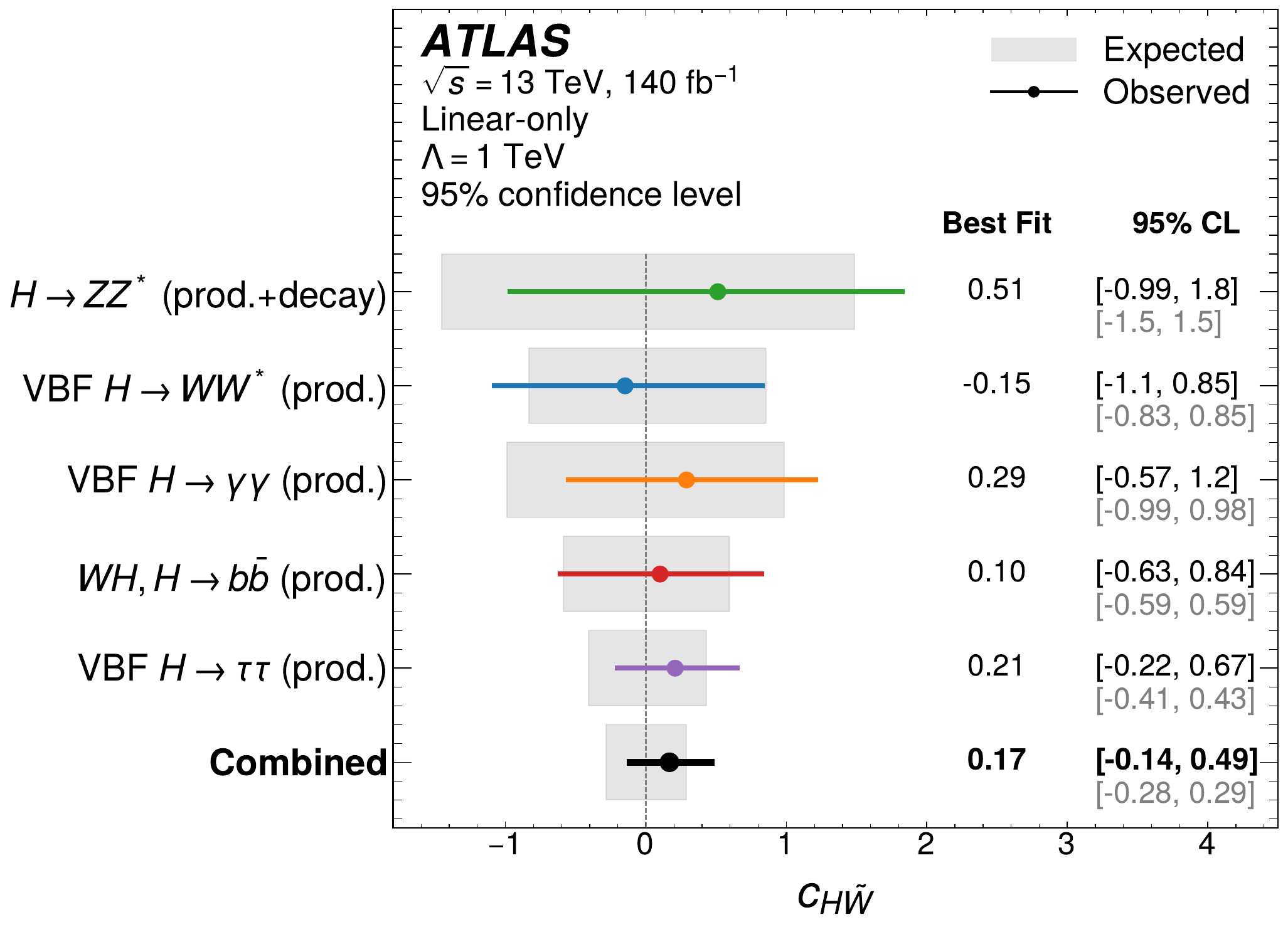}}
\subfloat[]{\includegraphics[width=0.46\linewidth]{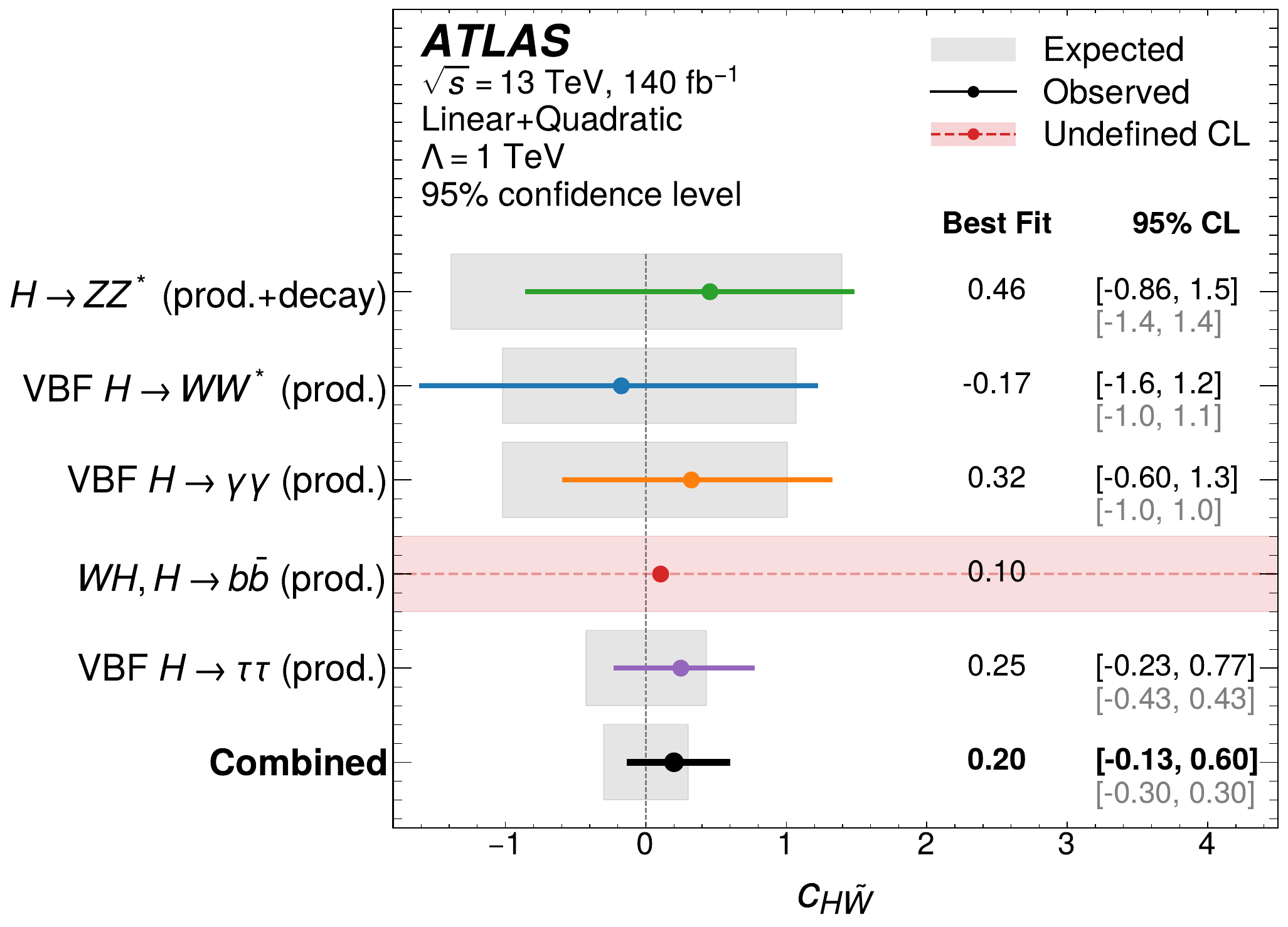}}
\caption{Best-fit values and 95\% CL intervals for \cHWtil~obtained in single-parameter fits for each of the individual analyses and the combined result, for both (a) linear-only and (b) linear and quadratic interpretations. For the \WHbb linear and quadratic result, the best-fit value is valid in the \cHWtil~range $[-1.5,1.5]$ and no 95\% CL interval is defined, as explained in the text.}
\label{fig:1d_and_3d_1poi_combined}
\end{figure}


\clearpage
\printbibliography

@article{SMEFTsim,
    author = "Brivio, Ilaria",
    title = "{SMEFTsim 3.0 \textemdash{} a practical guide}",
    eprint = "2012.11343",
    archivePrefix = "arXiv",
    primaryClass = "hep-ph",
    doi = "10.1007/JHEP04(2021)073",
    journal = "JHEP",
    volume = "04",
    pages = "073",
    year = "2021"
}

@article{Warsaw,
    author = "Grzadkowski, B. and Iskrzynski, M. and Misiak, M. and Rosiek, J.",
    title = "{Dimension-Six Terms in the Standard Model Lagrangian}",
    eprint = "1008.4884",
    archivePrefix = "arXiv",
    primaryClass = "hep-ph",
    reportNumber = "IFT-9-2010, TTP10-35",
    doi = "10.1007/JHEP10(2010)085",
    journal = "JHEP",
    volume = "10",
    pages = "085",
    year = "2010"
}

@article{STXS-Stage1.1,
	title={{Simplified template cross sections – Stage 1.1 and 1.2}},
	author={Nicolas Berger and Claudia Bertella and Matteo Bonanomi and Nihal Brahimi and Thomas P. Calvet and Milene Calvetti and Valerio Dao and Marco Delmastro and Michael Duehrssen-Debling and Paolo Francavilla and Yacine Haddad and Sarah Heim and Jelena Jovicevic and Oleh Kivernyk and Maria Moreno Llacer and Jonathon M. Langford and Changqiao Li and Giovanni Marchiori and Josh A. McFayden and Johannes K. L. Michel and Predrag Milenovic and Carlo E. Pandini and Edward Scott and Frank J. Tackmann and Kerstin Tackmann and Lorenzo Viliani and Meng Xiao and Hongtao Yang},
	journal={SciPost Phys. Comm. Rep.},
	pages={15},
	year={2026},
	publisher={SciPost},
	doi={10.21468/SciPostPhysCommRep.15},
  eprint = "1906.02754",
  archivePrefix = "arXiv",
  primaryClass = "hep-ph",
}

@article{STXS-LesHouches,
    author = "Badger S. and others",
    title = "{Les Houches 2015: Physics at TeV Colliders Standard Model Working Group Report}",
    booktitle = "{9th Les Houches Workshop on Physics at TeV Colliders}",
    eprint = "1605.04692",
    archivePrefix = "arXiv",
    primaryClass = "hep-ph",
    reportNumber = "FERMILAB-CONF-16-175-PPD-T",
    month = "5",
    year = "2016"
}

@article{GodboleHCP,
        archivePrefix = {arXiv},
        arxivId = {1409.5449},
        author = {Godbole, R. M. and Miller, D. J. and Mohan, K. A. and White, C. D.},
        doi = {10.1007/JHEP04(2015)103},
        eprint = {1409.5449},
        file = {:media/utilizador/Data/CERNBox/Articles/Jet substructure and probes of CP violation in Vh production - Godbole et al.pdf:pdf},
        journal = {JHEP},
        keywords = {Beyond Standard Model,CP violation,Hadron-Hadron Scattering,Higgs physics,Jet substructure},
        mendeley-groups = {Effective Field Theory/Phenomenology,MEAFP I},
        %month = {apr},
        number = {4},
        pages = {103},
        title = {{Jet substructure and probes of CP violation in Vh production}},
        volume = {2015},
        year = {2015}
}

@article{sally_cpv_wh,
    author = "Barrué, Ricardo and C. Muíño, Patricia and Dao, Valerio and Santos, Rui",
    title = "{Simulation-based inference in the search for CP violation in leptonic WH production}",
    eprint = "2308.02882",
    archivePrefix = "arXiv",
    primaryClass = "hep-ph",  
    doi = "10.1007/JHEP04(2024)014",
    journal = "JHEP",
    volume = "04",
    pages = "014",
    year = "2024"
}

@Article{Sakharov:1967dj,
  author  = {Sakharov, A. D},
  journal = {Sov.Phys.Usp.},
  title   = {{Violation of CP Invariance, C asymmetry, and baryon asymmetry of the universe}},
  year    = {1991},
  pages   = {392},
  volume  = {5},
  doi     = {10.1070/PU1991v034n05ABEH002497},
}

@Article{Cabibbo:1963yz,
  author  = {Cabibbo, Nicola},
  journal = {Phys. Rev. Lett.},
  title   = {{Unitary Symmetry and Leptonic Decays}},
  year    = {1963},
  pages   = {531--533},
  volume  = {10},
  doi     = {10.1103/PhysRevLett.10.531},
}

@Article{Kobayashi:1973fv,
  author       = {Kobayashi, Makoto and Maskawa, Toshihide},
  journal      = {Progress of Theoretical Physics},
  title        = {{CP-Violation in the Renormalizable Theory of Weak Interaction}},
  year         = {1973},
  pages        = {652--657},
  volume       = {49},
  doi          = {10.1143/PTP.49.652},
  reportnumber = {KUNS-242},
}

@Article{Brivio:2017btx,
  author        = {Brivio, Ilaria and Jiang, Yun and Trott, Michael},
  journal       = {JHEP},
  title         = {{The SMEFTsim package, theory and tools}},
  year          = {2017},
  pages         = {070},
  volume        = {12},
  archiveprefix = {arXiv},
  doi           = {10.1007/JHEP12(2017)070},
  eprint        = {1709.06492},
  primaryclass  = {hep-ph},
}

@article{Degrande:2016dqg,
    author = "Degrande, Celine and Fuks, Benjamin and Mawatari, Kentarou and Mimasu, Ken and Sanz, Veronica",
    title = "{Electroweak Higgs boson production in the standard model effective field theory beyond leading order in QCD}",
    eprint = "1609.04833",
    archivePrefix = "arXiv",
    primaryClass = "hep-ph",
    reportNumber = "IPPP-16-80, LPSC16198",
    doi = "10.1140/epjc/s10052-017-4793-x",
    journal = "Eur. Phys. J. C",
    volume = "77",
    number = "4",
    pages = "262",
    year = "2017"
}

@article{Atwood:1991ka,
    author = "Atwood, D. and Soni, A.",
    title = "{Analysis for magnetic moment and electric dipole moment form-factors of the top quark via $e^+ e^- \to t \bar{t}$}",
    reportNumber = "BNL-46892",
    doi = "10.1103/PhysRevD.45.2405",
    journal = "Phys. Rev. D",
    volume = "45",
    pages = "2405--2413",
    year = "1992"
}

@article{Davier:1992nw,
    author = "Davier, M. and Duflot, L. and Le Diberder, F. and Rougé, A.",
    title = "{The Optimal method for the measurement of tau polarization}",
    reportNumber = "LAL-92-73, X-LPNHE-92-22",
    doi = "10.1016/0370-2693(93)90101-M",
    journal = "Phys. Lett. B",
    volume = "306",
    pages = "411--417",
    year = "1993",
    url = "https://cds.cern.ch/record/246007"
}

@inproceedings{Diehl:1996wm,
    author = "Diehl, M. and Nachtmann, O.",
    title = "{Optimal observables for measuring three-gauge-boson couplings in $e^+ e^- \to W^+ W^-$}", 
    booktitle = "{Physics with $e^{+}e^{-}$ Linear Colliders (The European Working Groups 4 Feb - 1 Sep 1995: Session 2)}",
    eprint = "hep-ph/9603207",
    archivePrefix = "arXiv",
    reportNumber = "DAMTP-96-25",
    pages = "301--304",
    month = "3",
    year = "1996"
}

@article{Bahl:2022yrs,
    author = "Bahl, Henning and Fuchs, Elina and Heinemeyer, Sven and Katzy, Judith and Menen, Marco and Peters, Krisztian and Saimpert, Matthias and Weiglein, Georg",
    title = "{Constraining the ${\mathcal {C}}{\mathcal {P}}$ structure of Higgs-fermion couplings with a global LHC fit, the electron EDM and baryogenesis}",
    eprint = "2202.11753",
    archivePrefix = "arXiv",
    primaryClass = "hep-ph",
    reportNumber = "CERN-TH-2021-231, DESY-22-033, EFI-22-1, IFT--UAM/CSIC--21-148",
    doi = "10.1140/epjc/s10052-022-10528-1",
    journal = "Eur. Phys. J. C",
    volume = "82",
    number = "7",
    pages = "604",
    year = "2022"
}

@article{Buchmuller:1985jz,
    author = "Buchmuller, W. and Wyler, D.",
    title = "{Effective Lagrangian Analysis of New Interactions and Flavor Conservation}",
    reportNumber = "CERN-TH-4254/85",
    doi = "10.1016/0550-3213(86)90262-2",
    journal = "Nucl. Phys. B",
    volume = "268",
    pages = "621--653",
    year = "1986",
    url= "http://cds.cern.ch/record/163116"
}

@article{Brivio:2017vri,
    author = "Brivio, Ilaria and Trott, Michael",
    title = "{The Standard Model as an Effective Field Theory}",
    eprint = "1706.08945",
    archivePrefix = "arXiv",
    primaryClass = "hep-ph",
    doi = "10.1016/j.physrep.2018.11.002",
    journal = "Phys. Rept.",
    volume = "793",
    pages = "1--98",
    year = "2019"
}

@article{Sjostrand:2014zea,
      author         = "Sj{\"o}strand, Torbj{\"o}rn and Ask, Stefan and Christiansen,
                        Jesper R. and Corke, Richard and Desai, Nishita and Ilten,
                        Philip and Mrenna, Stephen and Prestel, Stefan and
                        Rasmussen, Christine O. and Skands, Peter Z.",
      title          = "{An introduction to PYTHIA 8.2}",
      journal        = "Comput. Phys. Commun.",
      volume         = "191",
      year           = "2015",
      pages          = "159",
      doi            = "10.1016/j.cpc.2015.01.024",
      eprint         = "1410.3012",
      archivePrefix  = "arXiv",
      primaryClass   = "hep-ph",
      reportNumber   = "LU-TP-14-36, MCNET-14-22, CERN-PH-TH-2014-190,
                        FERMILAB-PUB-14-316-CD, DESY-14-178, SLAC-PUB-16122,
                        --FERMILAB-PUB-14-316-CD",
      SLACcitation   = "%%CITATION = ARXIV:1410.3012;%%"
}

@Article{Alwall:2014hca,
      author         = "Alwall, J. and Frederix, R. and Frixione, S. and Hirschi,
                        V. and Maltoni, F. and Mattelaer, O. and Shao, H. -S. and
                        Stelzer, T. and Torrielli, P. and Zaro, M.",
      title          = "{The automated computation of tree-level and
                        next-to-leading order differential cross sections, and
                        their matching to parton shower simulations}",
      journal        = "JHEP",
      volume         = "07",
      year           = "2014",
      pages          = "079",
      doi            = "10.1007/JHEP07(2014)079",
      eprint         = "1405.0301",
      archivePrefix  = "arXiv",
      primaryClass   = "hep-ph",
      reportNumber   = "CERN-PH-TH-2014-064, CP3-14-18, LPN14-066, MCNET-14-09,
                        ZU-TH-14-14",
      SLACcitation   = "%%CITATION = ARXIV:1405.0301;%%"
}

@Article{Nason:2004rx,
      author         = "Nason, Paolo",
      title          = "{A new method for combining NLO QCD with shower Monte Carlo algorithms}",
      journal        = "JHEP",
      volume         = "11",
      pages          = "040",
      doi            = "10.1088/1126-6708/2004/11/040",
      year           = "2004",
      eprint         = "hep-ph/0409146",
      archivePrefix  = "arXiv",
}

@Article{Frixione:2007vw,
      author         = "Frixione, Stefano and Nason, Paolo and Oleari, Carlo",
      title          = "{Matching NLO QCD computations with parton shower
                        simulations: the POWHEG method}",
      journal        = "JHEP",
      volume         = "11",
      pages          = "070",
      doi            = "10.1088/1126-6708/2007/11/070",
      year           = "2007",
      eprint         = "0709.2092",
      archivePrefix  = "arXiv",
      primaryClass   = "hep-ph",
}

@Article{Alioli:2010xd,
      author         = "Alioli, Simone and Nason, Paolo and Oleari, Carlo and Re,
                        Emanuele",
      title          = "{A general framework for implementing NLO calculations in
                        shower Monte Carlo programs: the POWHEG BOX}",
      journal        = "JHEP",
      volume         = "06",
      pages          = "043",
      doi            = "10.1007/JHEP06(2010)043",
      year           = "2010",
      eprint         = "1002.2581",
      archivePrefix  = "arXiv",
      primaryClass   = "hep-ph",
}

@article{deFlorian:2016spz,
      author         = "de Florian, D. and others",
      title          = "{Handbook of LHC Higgs Cross Sections: 4. Deciphering the Nature of the Higgs Sector}",
      collaboration  = "LHC Higgs Cross Section Working Group",
      doi            = "10.23731/CYRM-2017-002",
      year           = "2017",
      eprint         = "1610.07922",
      archivePrefix  = "arXiv",
      primaryClass   = "hep-ph",
      reportNumber   = "FERMILAB-FN-1025-T, CERN-2017-002-M",
      SLACcitation   = "%%CITATION = ARXIV:1610.07922;%%"
}

@article{Hamilton:2013fea,
      author         = "Hamilton, Keith and Nason, Paolo and Re, Emanuele and
                        Zanderighi, Giulia",
      title          = "{NNLOPS simulation of Higgs boson production}",
      journal        = "JHEP",
      volume         = "10",
      year           = "2013",
      pages          = "222",
      doi            = "10.1007/JHEP10(2013)222",
      eprint         = "1309.0017",
      archivePrefix  = "arXiv",
      primaryClass   = "hep-ph",
      reportNumber   = "MCNET-13-11, CERN-PH-TH-2013-205, OUTP-13-18P",
      SLACcitation   = "%%CITATION = ARXIV:1309.0017;%%"
}

@article{Hamilton:2015nsa,
      author         = "Hamilton, Keith and Nason, Paolo and Zanderighi, Giulia",
      title          = "{Finite quark-mass effects in the NNLOPS POWHEG+MiNLO
                        Higgs generator}",
      journal        = "JHEP",
      volume         = "05",
      year           = "2015",
      pages          = "140",
      doi            = "10.1007/JHEP05(2015)140",
      eprint         = "1501.04637",
      archivePrefix  = "arXiv",
      primaryClass   = "hep-ph",
      reportNumber   = "CERN-PH-TH-2015-006",
      SLACcitation   = "%%CITATION = ARXIV:1501.04637;%%"
}

@article{Nason:2009ai,
      author         = "Nason, Paolo and Oleari, Carlo",
      title          = "{NLO Higgs boson production via vector-boson fusion
                        matched with shower in POWHEG}",
      journal        = "JHEP",
      volume         = "02",
      year           = "2010",
      pages          = "037",
      doi            = "10.1007/JHEP02(2010)037",
      eprint         = "0911.5299",
      archivePrefix  = "arXiv",
      primaryClass   = "hep-ph",
      SLACcitation   = "%%CITATION = ARXIV:0911.5299;%%"
}

@Article{PERF-2007-01,
    author         = "{ATLAS Collaboration}",
    title          = "{The ATLAS Experiment at the CERN Large Hadron Collider}",
    journal        = "JINST",
    volume         = "3",
    year           = "2008",
    pages          = "S08003",
    doi            = "10.1088/1748-0221/3/08/S08003",
    primaryClass   = "hep-ex",
}

@Article{HIGG-2012-27,
    author         = "{ATLAS Collaboration}",
    title          = "{Observation of a new particle in the search for the Standard Model Higgs boson with the ATLAS detector at the LHC}",
    journal        = "Phys. Lett. B",
    volume         = "716",
    year           = "2012",
    pages          = "1",
    doi            = "10.1016/j.physletb.2012.08.020",
    reportNumber   = "CERN-PH-EP-2012-218",
    eprint         = "1207.7214",
    archivePrefix  = "arXiv",
    primaryClass   = "hep-ex",
}

@Article{HIGG-2017-13,
    author         = "{ATLAS Collaboration}",
    title          = "{Constraints on Higgs boson properties using \(WW^{*}(\rightarrow e\nu\mu\nu) jj\) production in \(36.1\,\text{fb}^{-1}\) of \(\sqrt{s} = 13\,\text{TeV}\) \(pp\) collisions with the ATLAS detector}",
    journal        = "Eur. Phys. J. C",
    volume         = "82",
    year           = "2022",
    pages          = "622",
    doi            = "10.1140/epjc/s10052-022-10366-1",
    reportNumber   = "CERN-EP-2021-096",
    eprint         = "2109.13808",
    archivePrefix  = "arXiv",
    primaryClass   = "hep-ex",
}

@Article{HIGG-2018-14,
    author         = "{ATLAS Collaboration}",
    title          = "{Test of CP invariance in vector-boson fusion production of the Higgs boson in the \(H\rightarrow\tau\tau\) channel in proton--proton collisions at \(\sqrt{s} = 13\,\text{TeV}\) with the ATLAS detector}",
    journal        = "Phys. Lett. B",
    volume         = "805",
    year           = "2020",
    pages          = "135426",
    doi            = "10.1016/j.physletb.2020.135426",
    reportNumber   = "CERN-EP-2020-009",
    eprint         = "2002.05315",
    archivePrefix  = "arXiv",
    primaryClass   = "hep-ex",
}

@Article{HIGG-2018-30,
    author         = "{ATLAS Collaboration}",
    title          = "{Test of CP-invariance of the Higgs boson in vector-boson fusion production and its decay into four leptons}",
    journal        = "JHEP",
    volume         = "05",
    year           = "2024",
    pages          = "105",
    doi            = "10.1007/JHEP05(2024)105",
    reportNumber   = "CERN-EP-2023-030",
    eprint         = "2304.09612",
    archivePrefix  = "arXiv",
    primaryClass   = "hep-ex",
}

@Article{GENR-2019-02,
    author         = "{ATLAS Collaboration}",
    title          = "{The ATLAS experiment at the CERN Large Hadron Collider: a description of the detector configuration for Run~3}",
    journal        = "JINST",
    volume         = "19",
    year           = "2024",
    pages          = "P05063",
    doi            = "10.1088/1748-0221/19/05/P05063",
    reportNumber   = "CERN-EP-2022-259",
    eprint         = "2305.16623",
    archivePrefix  = "arXiv",
    primaryClass   = "physics.ins-det",
}

@Article{HIGG-2019-01,
    author         = "{ATLAS Collaboration}",
    title          = "{\(CP\) Properties of Higgs Boson Interactions with Top Quarks in the \(t\bar{t}H\) and \(tH\) Processes Using \(H \rightarrow \gamma\gamma\) with the ATLAS Detector}",
    journal        = "Phys. Rev. Lett.",
    volume         = "125",
    year           = "2020",
    pages          = "061802",
    doi            = "10.1103/PhysRevLett.125.061802",
    reportNumber   = "CERN-EP-2020-046",
    eprint         = "2004.04545",
    archivePrefix  = "arXiv",
    primaryClass   = "hep-ex",
}

@Article{HIGG-2019-13,
    author         = "{ATLAS Collaboration}",
    title          = "{Measurements of the Higgs boson inclusive and differential fiducial cross-sections in the diphoton decay channel with \(pp\) collisions at \(\sqrt{s} = 13\,\text{TeV}\) with the ATLAS detector}",
    journal        = "JHEP",
    volume         = "08",
    year           = "2022",
    pages          = "027",
    doi            = "10.1007/JHEP08(2022)027",
    reportNumber   = "CERN-EP-2021-227",
    eprint         = "2202.00487",
    archivePrefix  = "arXiv",
    primaryClass   = "hep-ex",
}

@Article{HIGG-2020-08,
    author         = "{ATLAS Collaboration}",
    title          = "{Test of \(CP\) Invariance in Higgs Boson Vector-Boson-Fusion Production Using the \(H \rightarrow \gamma\gamma\) Channel with the ATLAS Detector}",
    journal        = "Phys. Rev. Lett.",
    volume         = "131",
    year           = "2023",
    pages          = "061802",
    doi            = "10.1103/PhysRevLett.131.061802",
    reportNumber   = "CERN-EP-2022-134",
    eprint         = "2208.02338",
    archivePrefix  = "arXiv",
    primaryClass   = "hep-ex",
}

@Article{HIGG-2020-20,
    author         = "{ATLAS Collaboration}",
    title          = "{Measurements of \(WH\) and \(ZH\) production with Higgs boson decays into bottom quarks and direct constraints on the charm Yukawa coupling in \(13\,\text{TeV}\) \(pp\) collisions with the ATLAS detector}",
    journal        = "JHEP",
    volume         = "04",
    year           = "2025",
    pages          = "075",
    doi            = "10.1007/JHEP04(2025)075",
    reportNumber   = "CERN-EP-2024-237",
    eprint         = "2410.19611",
    archivePrefix  = "arXiv",
    primaryClass   = "hep-ex",
}

@Article{SOFT-2022-02,
    author         = "{ATLAS Collaboration}",
    title          = "{Software and computing for Run~3 of the ATLAS experiment at the LHC}",
    journal        = "Eur. Phys. J. C",
    volume         = "85",
    year           = "2025",
    pages          = "234",
    doi            = "10.1140/epjc/s10052-024-13701-w",
    reportNumber   = "CERN-EP-2024-100",
    eprint         = "2404.06335",
    archivePrefix  = "arXiv",
    primaryClass   = "hep-ex",
    related        = "SOFT-2022-02-err",
    relatedstring  = "Erratum:",
}

@Article{HIGP-2024-07,
    author         = "{ATLAS Collaboration}",
    title          = "{Measurements of Higgs boson production via gluon--gluon fusion and vector-boson fusion using \(H \rightarrow WW^{*} \rightarrow \ell\nu\ell\nu\) decays in \(pp\) collisions with the ATLAS detector and their effective field theory interpretations}",
    journal        = "Eur. Phys. J. C",
    volume         = "85",
    year           = "2025",
    pages          = "1403",
    doi            = "10.1140/epjc/s10052-025-14761-2",
    reportNumber   = "CERN-EP-2025-054",
    eprint         = "2504.07686",
    archivePrefix  = "arXiv",
    primaryClass   = "hep-ex",
}

@Article{HIGP-2024-09,
    author         = "{ATLAS Collaboration}",
    title          = "{Probing the Higgs boson CP properties in vector-boson fusion production in the \(H \rightarrow \tau^{+}\tau^{-}\) channel with the ATLAS detector}",
    journal        = "JHEP",
    volume         = "10",
    year           = "2025",
    pages          = "092",
    doi            = "10.1007/JHEP10(2025)092",
    reportNumber   = "CERN-EP-2025-127",
    eprint         = "2506.19395",
    archivePrefix  = "arXiv",
    primaryClass   = "hep-ex",
}

@Article{HIGP-2024-26,
    author         = "{ATLAS Collaboration}",
    title          = "{Combination of measurements of CP properties of Higgs boson interactions with vector bosons using proton--proton collisions at \(\sqrt{s} = 13\,\text{TeV}\) with the ATLAS detector}",
    year           = "2026",
    reportNumber   = "CERN-EP-2026-071",
    eprint         = "2603.20117",
    archivePrefix  = "arXiv",
    primaryClass   = "hep-ex",
}

@Booklet{ATL-SOFT-PUB-2026-001,
    author         = "{ATLAS Collaboration}",
    title          = "{ATLAS Computing Acknowledgements}",
    howpublished   = "{ATL-SOFT-PUB-2026-001}",
    url            = "https://cds.cern.ch/record/2952666",
    year           = "2026",
}

@Article{CMS-HIG-12-028,
    author         = "{CMS Collaboration}",
    title          = "{Observation of a new boson at a mass of 125 GeV with the CMS experiment at the LHC}",
    journal        = "Phys. Lett. B",
    volume         = "716",
    year           = "2012",
    pages          = "30",
    doi            = "10.1016/j.physletb.2012.08.021",
    reportNumber   = "CERN-PH-EP-2012-220",
    eprint         = "1207.7235",
    archivePrefix  = "arXiv",
    primaryClass   = "hep-ex",
}

@Article{CMS-HIG-17-034,
    author         = "{CMS Collaboration}",
    title          = "{Constraints on anomalous \(HVV\) couplings from the production of Higgs bosons decaying to \(\tau\) lepton pairs}",
    journal        = "Phys. Rev. D",
    volume         = "100",
    year           = "2019",
    pages          = "112002",
    doi            = "10.1103/PhysRevD.100.112002",
    reportNumber   = "CERN-EP-2019-029",
    eprint         = "1903.06973",
    archivePrefix  = "arXiv",
    primaryClass   = "hep-ex",
}

@Article{CMS-HIG-19-009,
    author         = "{CMS Collaboration}",
    title          = "{Constraints on anomalous Higgs boson couplings to vector bosons and fermions in its production and decay using the four-lepton final state}",
    journal        = "Phys. Rev. D",
    volume         = "104",
    year           = "2021",
    pages          = "052004",
    doi            = "10.1103/PhysRevD.104.052004",
    reportNumber   = "CERN-EP-2021-054",
    eprint         = "2104.12152",
    archivePrefix  = "arXiv",
    primaryClass   = "hep-ex",
}

@Article{CMS-HIG-19-013,
    author         = "{CMS Collaboration}",
    title          = "{Measurements of \(t\bar{t}H\) Production and the \(CP\) Structure of the Yukawa Interaction between the Higgs Boson and Top Quark in the Diphoton Decay Channel}",
    journal        = "Phys. Rev. Lett.",
    volume         = "125",
    year           = "2020",
    pages          = "061801",
    doi            = "10.1103/PhysRevLett.125.061801",
    reportNumber   = "CERN-EP-2020-028",
    eprint         = "2003.10866",
    archivePrefix  = "arXiv",
    primaryClass   = "hep-ex",
}

@Article{CMS-HIG-20-006,
    author         = "{CMS Collaboration}",
    title          = "{Analysis of the CP structure of the Yukawa coupling between the Higgs boson and \(\tau\) leptons in proton--proton collisions at \(\sqrt{s} = 13\,\text{TeV}\)}",
    journal        = "JHEP",
    volume         = "06",
    year           = "2022",
    pages          = "012",
    doi            = "10.1007/JHEP06(2022)012",
    reportNumber   = "CERN-EP-2021-189",
    eprint         = "2110.04836",
    archivePrefix  = "arXiv",
    primaryClass   = "hep-ex",
}

@Article{CMS-HIG-20-007,
    author         = "{CMS Collaboration}",
    title          = "{Constraints on anomalous Higgs boson couplings to vector bosons and fermions from the production of Higgs bosons using the \(\tau\tau\) final state}",
    journal        = "Phys. Rev. D",
    volume         = "108",
    year           = "2023",
    pages          = "032013",
    doi            = "10.1103/PhysRevD.108.032013",
    reportNumber   = "CERN-EP-2022-021",
    eprint         = "2205.05120",
    archivePrefix  = "arXiv",
    primaryClass   = "hep-ex",
}

@Article{CMS-HIG-22-008,
    author         = "{CMS Collaboration}",
    title          = "{Constraints on anomalous Higgs boson couplings from its production and decay using the \(WW\) channel in proton--proton collisions at \(\sqrt{s} = 13\,\text{TeV}\)}",
    journal        = "Eur. Phys. J. C",
    volume         = "84",
    year           = "2024",
    pages          = "779",
    doi            = "10.1140/epjc/s10052-024-12925-0",
    reportNumber   = "CERN-EP-2024-035",
    eprint         = "2403.00657",
    archivePrefix  = "arXiv",
    primaryClass   = "hep-ex",
}
\clearpage
 
\begin{flushleft}
\hypersetup{urlcolor=black}
{\Large The ATLAS Collaboration}

\bigskip

\AtlasOrcid[0000-0002-6665-4934]{G.~Aad}$^\textrm{\scriptsize 102}$,
\AtlasOrcid[0000-0001-7616-1554]{E.~Aakvaag}$^\textrm{\scriptsize 17}$,
\AtlasOrcid[0000-0002-5888-2734]{B.~Abbott}$^\textrm{\scriptsize 121}$,
\AtlasOrcid[0000-0002-0287-5869]{S.~Abdelhameed}$^\textrm{\scriptsize 83b}$,
\AtlasOrcid[0000-0002-1002-1652]{K.~Abeling}$^\textrm{\scriptsize 54}$,
\AtlasOrcid[0000-0001-5763-2760]{N.J.~Abicht}$^\textrm{\scriptsize 48}$,
\AtlasOrcid[0000-0002-8496-9294]{S.H.~Abidi}$^\textrm{\scriptsize 30}$,
\AtlasOrcid[0009-0003-6578-220X]{M.~Aboelela}$^\textrm{\scriptsize 44}$,
\AtlasOrcid[0000-0002-9987-2292]{A.~Aboulhorma}$^\textrm{\scriptsize 36e}$,
\AtlasOrcid[0000-0001-5329-6640]{H.~Abramowicz}$^\textrm{\scriptsize 154}$
% ...
\bigskip
\\

% ...
$^{17}$Department for Physics and Technology, University of Bergen, Bergen; Norway.\\
% ...
$^{30}$Physics Department, Brookhaven National Laboratory, Upton NY; United States of America.\\
% ...
$^{36}$$^{(a)}$Facult\'e des Sciences Ain Chock, Universit\'e Hassan II de Casablanca;$^{(b)}$Facult\'{e} des Sciences, Universit\'{e} Ibn-Tofail, K\'{e}nitra;$^{(c)}$Facult\'e des Sciences Semlalia, Universit\'e Cadi Ayyad, LPHEA-Marrakech;$^{(d)}$LPMR, Facult\'e des Sciences, Universit\'e Mohamed Premier, Oujda;$^{(e)}$Facult\'e des sciences, Universit\'e Mohammed V, Rabat;$^{(f)}$Institute of Applied Physics, Mohammed VI Polytechnic University, Ben Guerir; Morocco.\\
% ...
$^{44}$Physics Department, Southern Methodist University, Dallas TX; United States of America.\\
% ...
$^{48}$Fakult\"{a}t Physik, Technische Universit{\"a}t Dortmund, Dortmund; Germany.\\
% ...
$^{54}$II. Physikalisches Institut, Georg-August-Universit\"{a}t G\"ottingen, G\"ottingen; Germany.\\
% ...
$^{83}$$^{(a)}$Khalifa University of Science and Technology, Abu Dhabi;$^{(b)}$New York University Abu Dhabi, Abu Dhabi;$^{(c)}$United Arab Emirates University, Al Ain;$^{(d)}$University of Sharjah, Sharjah; United Arab Emirates.\\
% ...
$^{102}$CPPM, Aix-Marseille Universit\'e, CNRS/IN2P3, Marseille; France.\\
% ...
$^{121}$Homer L. Dodge Department of Physics and Astronomy, University of Oklahoma, Norman OK; United States of America.\\
% ...
$^{154}$Raymond and Beverly Sackler School of Physics and Astronomy, Tel Aviv University, Tel Aviv; Israel.\\
% ...
$^{a}$ Also at Affiliated with an institute formerly covered by a cooperation agreement with CERN.\\
$^{b}$ Also at An-Najah National University, Nablus; Palestine.\\
$^{c}$ Also at Borough of Manhattan Community College, City University of New York, New York NY; United States of America.\\
$^{d}$ Also at Center for Interdisciplinary Research and Innovation (CIRI-AUTH), Thessaloniki; Greece.\\
$^{e}$ Also at Centre of Physics of the Universities of Minho and Porto (CF-UM-UP); Portugal.\\
$^{f}$ Also at CERN, Geneva; Switzerland.\\
% ...

\end{flushleft}


%

%
%

%

%
%
%
%

\end{document}